%% file: charged_mediators.tex
\documentclass[11pt]{article}
\usepackage{graphicx}
\usepackage[margin=1.25in]{geometry}
\usepackage[usenames,dvipsnames]{color}
\usepackage{url}
\usepackage[colorlinks = true,
            linkcolor = blue,
            urlcolor  = blue,
            citecolor = blue,
            anchorcolor = blue]{hyperref}

\usepackage{amsmath}
\usepackage{amssymb}

\newcommand\pubnumber{HRI-RECAPP-2022-005}
\newcommand\pubdate{\today}

\def\Title#1{\begin{center} {\LARGE #1 } \end{center}}
\def\Author#1{\begin{center}{ \sc #1} \end{center}}
\def\Address#1{\begin{center}{ \it #1} \end{center}}

\newcommand\pubblock{\rightline{\begin{tabular}{l} \pubnumber\\
         \pubdate \end{tabular}}}
\newenvironment{Abstract}{\begin{quotation} \begin{center}
                       ABSTRACT
     \end{center}\bigskip  }{\end{quotation}}

\def\Acknowledgements{\bigskip  \bigskip \begin{center} \begin{large}
             \bf ACKNOWLEDGEMENTS \end{large}\end{center}}
             
             \def\be{\begin{equation}}
\def\ee{\end{equation}}
\def\bea{\begin{eqnarray}}
\def\eea{\end{eqnarray}}

\def\tev{\, {\rm TeV}}
\def\gev{\, {\rm GeV}}

\newcommand{\sigmaSI}{\sigma_{\rm SI}}

\newcommand{\gsim}{\lower.7ex\hbox{$\;\stackrel{\textstyle>}{\sim}\;$}}
\newcommand{\lsim}{\lower.7ex\hbox{$\;\stackrel{\textstyle<}{\sim}\;$}}

\newcommand{\pb}{\rm pb}

\input workshopsymbols.tex

\newcommand\snowmass{\begin{center}\rule[-0.2in]{\hsize}{0.01in}\\\rule{\hsize}{0.01in}\\
\vskip 0.1in Submitted to the  Proceedings of the US Community Study\\ 
on the Future of Particle Physics (Snowmass 2021)\\ 
\rule{\hsize}{0.01in}\\\rule[+0.2in]{\hsize}{0.01in} \end{center}}

\DeclareUnicodeCharacter{2212}{-}

\begin{document}

\pubblock

\Title{Simplified dark matter models with charged mediators}

\bigskip 

\Author{Tathagata Ghosh$^a$, Chris Kelso$^b$, Jason Kumar$^c$, Pearl Sandick$^d$ \\ and Patrick Stengel$^{e,f,g}$}

\medskip

\Address{
         $\,^a$ Regional Centre for Accelerator-based Particle Physics,
         Harish-Chandra Research Institute,
         A CI of Homi Bhabha National Institute,
         Chhatnag Road, Jhusi, Prayagraj 211019, India \\
         $\,^b$ Department of Physics, University of North Florida, Jacksonville, FL 32224, USA \\ 
         $\,^c$ Department of Physics \& Astronomy, University of Hawaii, Honolulu, HI 96822, USA \\
         $\,^d$ Department of Physics and Astronomy, University of Utah, Salt Lake City, UT 84112, USA \\
         $\,^e$ Scuola Internazionale Superiore di Studi Avanzati (SISSA), via Bonomea 265, 34136 Trieste, Italy \\ 
         $\,^f$ INFN, Sezione di Trieste, via Valerio 2, 34127 Trieste, Italy \\
         $\,^g$ Institute for Fundamental Physics of the Universe (IFPU), via Beirut 2, 34151 Trieste, Italy 
         }

\medskip

 \begin{Abstract}
\noindent We review simplified models in which a singlet Majorana dark matter candidate couples to Standard Model (SM) fermions through interactions mediated by scalar fermion partners. We summarize the two primary production mechanisms in these scenarios: dark matter annihilation mediated by first or second generation scalar fermion partners with significant left-right chiral mixing and co-annihilation with scalar fermion partners nearly degenerate in mass with the dark matter. We then highlight the most interesting phenomenological aspects of charged mediator models relevant for current and future searches for new physics. We describe precision measurements of SM fermion dipole moments, including models with scalar muon partners that can account for $g_\mu-2$. We discuss new search strategies for charged mediators at the LHC and the projected sensitivity of future lepton colliders. We summarize constraints from direct detection and demonstrate how next generation experiments might probe QCD-charged mediators at mass scales beyond the sensitivity of the LHC. We also review the prospects for indirect detection of models with scalar lepton partners, focusing on the sensitivity of gamma-ray searches to internal bremsstrahlung emission.
\end{Abstract}

\snowmass

\def\thefootnote{\fnsymbol{footnote}}
\setcounter{footnote}{0}

\section{Introduction}

We consider here a general and well-motivated scenario for new physics beyond the Standard Model (SM), in which a gauge-singlet Majorana fermion dark matter (DM) candidate couples to Standard Model fermions through charged scalar mediators. The most commonly studied implementation of this scenario arises in the minimal supersymmetric Standard Model (MSSM), in the case where the lightest supersymmetric particle is the bino, which couples to SM fermions through interactions mediated by sfermions. But this is a much more general scenario which exhibits an interesting suite of phenomenology, especially when the assumption of minimal flavor violation (MFV), common to MSSM scenarios, is lifted.

Perhaps the first notable aspect of this scenario is that it allows for new particles coupled to the Standard Model, with masses in the ${\cal O}(100\gev)$ range, which are not ruled out by tight constraints from the LHC.  Indeed, if the dark matter couples to leptons, with a mass splitting from the mediators of ${\cal O}(40\gev)$, then the large electroweak backgrounds (BG) lead to LHC bounds which show little improvement over LEP.

There are several other interesting features of this scenario which are relevant for cosmology and experiment:
\begin{itemize}

\item{If dark matter couples to leptons, then the relic density of a ${\cal O}(100\gev)$  thermal dark matter candidate can be appropriately depleted by $s$- or $p$-wave annihilation~\cite{Fukushima:2014yia}. This ``Incredible Bulk" scenario is a generalization of the usual bulk region considered in constrained MSSM (CMSSM) scenarios, for which the assumptions of MFV and small $CP$-violation are weakened.}

\item{If the mass splitting between the the dark matter and the lightest charged mediator is small, then the charged mediator may also be abundant at the time of dark matter thermal freeze out.  In this case, co-annihilation can sufficiently deplete the dark matter density~\cite{Davidson:2017gxx,Acuna:2021rbg}.}

\item{Although dark matter annihilation via internal bremsstrahlung processes are unlikely to play a dominant role in depleting the relic density at early times, it can play an important role in current indirect detection searches~\cite{Kumar:2016cum}.}

\item{Dark matter-nucleon scattering at direct detection experiments can be mediated by exchange of charged mediators~\cite{Kelso:2014qja,Sandick:2016zut}. If the mass splitting between the dark matter and the charged mediator is small, then the scattering rate can be strongly enhanced, giving upcoming direct detection experiments a mass reach which could rival the LHC for QCD-charged mediators.}

\item{The coupling of dark matter to muons will yield corrections to the magnetic and electric dipole moments of the muon~\cite{Fukushima:2013efa}. This provides tight constraints on this scenario, but also 
provides a window for explaining the anomalies in the $g_\mu-2$ data.}

\item{Although current LHC searches for dilepton production associated with missing transverse energy have yielded only limited improvement over LEP for some of the parameter space of interest, new search strategies at the LHC~\cite{Dutta:2017nqv}, and searches at a future high-energy lepton collider~\cite{Baum:2020gjj}, may be able to probe this scenario.}

\end{itemize}

The plan for this white paper is as follows. We outline the particle content and couplings for the charged mediator models of interest in Sec.~\ref{sec:model}. In Sec.~\ref{sec:relic}, we describe the production mechanisms relevant for simplified dark matter models with charged mediators. Constraints from current experiments and the sensitivity of future experiments are summarized in Sec.~\ref{sec:pheno}. We conclude with an overview of charged mediator models discussed here and briefly comment on future work in Sec.~\ref{sec:con}. 

\section{Charged mediator models} \label{sec:model}

We consider a simplified model with Majorana singlet DM, $\s{B}$, coupled to SM fermions, $f=\{\ell,q\}$, through the interactions~\footnote{The left chiral scalar partner of the neutrino similarly couples DM to the SM neutrino, $f=\nu$, only through the first term of Eq.~\leqn{eq:Lint}. For the work summarized in this white paper, the only role the scalar partner of the neutrino potentially plays is in the calculation of the relic density when the effects of co-annihilation are relevant~\cite{Acuna:2021rbg}.}
\begin{equation}
    \L_{\rm int} \ni \lambda_L \s{f}^*_L\s{B} P_L f + \lambda_R \s{f}^*_R \s{B} P_R f + c.c.  \, ,
    \label{eq:Lint}
\end{equation}
where $\s{f}_{L,R}$ are the left (right) chiral scalar partners of the SM fermions and $P_{L,R}$ are the left (right) chiral projectors. Gauge-invariance implies that the $\tilde f_{L,R}$ have the same gauge quantum numbers as $f_{L,R}$, and in particular, that $\tilde f_L$ is an $SU(2)_L$ doublet. If the $\tilde f$ and $\tilde B$ are charged under an unbroken $Z_2$ symmetry, then the lightest of the mass eigenstates (which we assume to be $\tilde B$) is a dark matter candidate.

The most general Yukawa couplings $\lambda_{L,R}$ allow for a $CP$-violating phase, $\phi$, and can be written as
\begin{equation} \label{eq:couplings}
    \lambda_L = |\lambda_L| \, e^{\imath \phi / 2} , \quad \quad
    \lambda_R = |\lambda_R| \, e^{- \imath \phi / 2} .
\end{equation}
In principle, the Yukawa couplings can take any value consistent with perturbative unitarity (\ie $\, |\lambda_{L,R}| \lsim \sqrt{4 \pi}$). However, as a benchmark we often refer to the bino-fermion-sfermion couplings of the MSSM, $|\lambda_{L,R}| = \sqrt{2} g |Y_{L,R}|$, where $g$ is the weak hypercharge coupling of the SM and $Y_{L,R}$ are the hypercharges of the associated SM fermions. As discussed below, the $CP$-violating phase can become relevant when reconciling $g_\mu - 2$ with the calculation of the DM relic density in models with scalar lepton partners.

The interactions described above are defined in terms of the chiral eigenstates of the scalar mediators. The mass eigenstates are related to the respective chiral eigenstates of each mediator through the mixing matrix 
\begin{equation}
\begin{pmatrix}
\s{f}_1 \\
\s{f}_2 
\end{pmatrix} 
=
\begin{pmatrix}
\cos \alpha & - \sin \alpha \\
\sin \alpha & \cos \alpha
\end{pmatrix}
\begin{pmatrix}
\s{f}_L \\
\s{f}_R 
\end{pmatrix} \, ,
\end{equation}
where $\alpha$ is the left-right mixing angle. As we discuss below, significant left-right mixing can be important for both the production of DM and its detection. For typical realizations of the MSSM, which assume minimal flavor violation (MFV), such mixing is usually small for first and second generation scalar mediators. Relaxing the assumption of MFV to allow for intragenerational mixing thus opens up a broad parameter space of viable DM models with interesting phenomenology, including signatures at direct detection experiments for models with QCD-charged mediators and accounting for $g_\mu - 2$ in models with scalar lepton partners. As described in Sec.~\ref{sec:dip}, constraints from electroweak precision measurements of, for example, fermion dipole moments can become relevant when allowing for intragenerational mixing. It is important to distinguish between this relatively limited deviation from MFV and even more general scenarios which allow for mixing between generations. The latter are much more severely constrained and, for simplicity, we to not consider models with intergenerational mixing in this white paper.

In addition to the interactions described by Eq.~\leqn{eq:Lint}, there are a large number of possible gauge-invariant and renormalizable terms which can couple the scalar mediators to SM gauge bosons and the Higgs sector. While such couplings do not impact the phenomenological signatures of charged mediator models, the associated interactions can significantly affect  the calculation of the DM relic density in the co-annihilation scenarios described in Sec.~\ref{sec:coann}. For QCD-charged mediators, the couplings to gluons can allow for the process which provides for the largest contribution to the annihilation rate, $\s{q}^*\s{q} \to g g$, in models which satisfy the DM relic density. Similarly, the scalar lepton partner couplings to electroweak gauge bosons and the Higgs can open up annihilation channels such as $\s{\ell}^*\s{\ell} \to W^+ W^-$ and $\s{\ell}^*\s{\ell} \to h h$, which can be the dominant processes that deplete the relic density. 

The various charged mediator models discussed in this paper can be categorized by SM charges of the mediators and the DM production mechanism in the early Universe. As we describe in Sec.~\ref{sec:pheno}, phenomenological signatures can differ significantly between models with scalar lepton partners and those with QCD-charged mediators. In principle, a more extended theoretical framework such as the MSSM may have both leptonic and QCD-charged mediators (borrowing from the nomenclature of the MSSM, hereafter referred to as sleptons and squarks, respectively). However, in this paper we focus on simplified models which are minimal extensions to the SM and only involve either sleptons or squarks mediating interactions between the Majorana singlet DM candidate (hereafter refereed to as the bino) and the associated SM fermions. 

\section{Relic Density} \label{sec:relic}

Irrespective of the SM charges of the mediators for a given model, the primary mechanism for DM production in such models can be described by either bino annihilation or co-annihilation involving the charged mediators. In this section we summarize the features most relevant for the production of DM in charged mediator models. Several more detailed equations are discussed in Appendix~\ref{app:relic}, while complete descriptions of the Incredible Bulk and co-annihilation scenarios can be be found in Refs.~\cite{Fukushima:2014yia} and~\cite{Davidson:2017gxx,Acuna:2021rbg}, respectively.

\subsection{Incredible Bulk} \label{sec:bulk}

The most extensively studied realizations of SUSY models include the constrained MSSM (CMSSM), which makes several assumptions based on theoretical and phenomenological considerations to simplify the rather vast parameter space of the MSSM. The low energy spectra in the ``bulk region" of CMSSM parameter space are typically characterized by a bino-like DM candidate, along with relatively light sleptons and squarks. Given the constraints on scalar mediator masses discussed in Sec.~\ref{sec:coll}, the bino annihilation rate is usually too small to satisfy the DM relic density in the bulk region without either a small admixture of the bino with the fermionic partners of the Higgs bosons (\ie~Higgsinos) or a spectrum which is sufficiently compressed such that the effects of co-annihilation become important. We discuss particular examples of the latter in Sec.~\ref{sec:coann}, and we do not consider the former in our simplified model with DM assumed to be a SM singlet.

\begin{figure}
\begin{center}
\includegraphics[width=0.5\hsize]{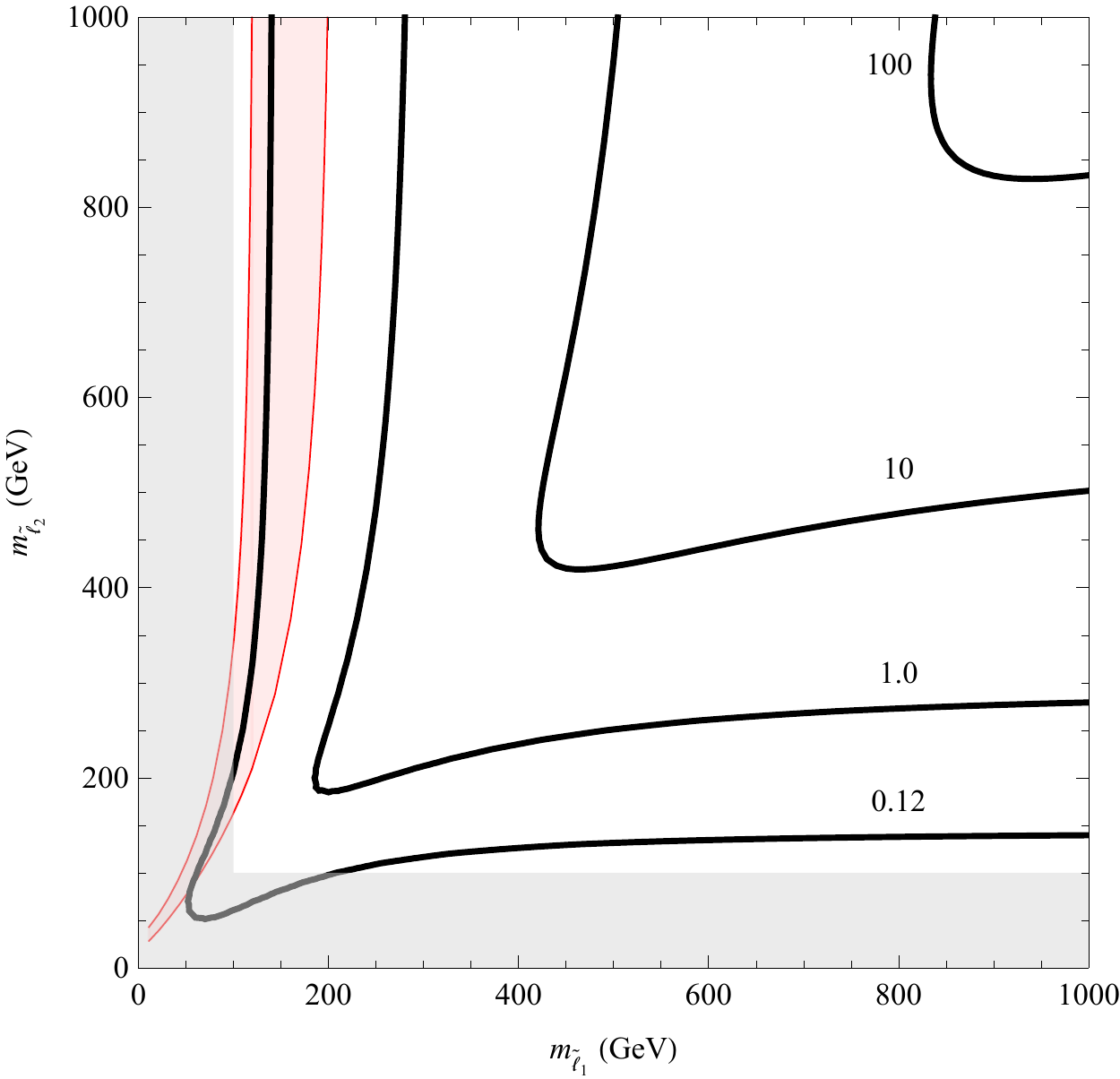}
\end{center}
\caption{Relic density contours as a function of slepton masses $m_{\s{\ell}_{1,2}}$ for fixed $m_{\s{B}} = 100 \, \GeV$ and $\alpha = \pi/4 + 0.02$. Red region indicates parameter space where model can satisfy $g_\mu - 2$, assuming a $CP$-violating phase $\phi = \pi/2 - 0.04$. Appears as Fig.~5 in Ref.~\cite{Fukushima:2014yia}.}
\label{fig:bulkrelic}
\end{figure}

Alternatively, the rate of bino annihilation in the bulk region of the CMSSM can be sufficiently large if there is a sizeable left-right mixing angle for the charged mediators coupling the bino to the SM fermions through the interactions described by Eq.~\leqn{eq:Lint}. For the corresponding $t$- and $u$-channel diagrams contributing to the process $\s{B} \s{B} \to \bar{f} f$, we can expand the annihilation cross section in powers of the thermal freeze out temperature $T_f$,
\begin{equation} \label{eqn:annx}
    \VEV{\sigma_{\s{B}\s{B}} v} \sim c_0 + c_1 \left(\frac{T_f}{m_{\s{B}}} \right) \, ,
\end{equation}
where the freeze out temperature is approximately $T_f \sim m_{\s{B}} / 25$. The coefficients $c_0$ and $c_1$ are defined in terms of the Lagrangian parameters in Eqs.~\leqn{eq:c0} and~\leqn{eq:c1} in Appendix~\ref{app:bulk}. For any particular model, the relic density is approximately given by
\begin{equation}
    \Omega_{\s{B}} h^2 \sim 0.1 \left(\frac{1 \, \pb}{\VEV{\sigma_{\s{B}\s{B}} v}} \right) \, .
\end{equation}
In this scenario, the DM relic density can be satisfied for $m_{\s{B}} \simeq \O(100) \, \GeV$ and $m_{\s{f}_1} \lesssim 200 \, \GeV$ given a sizeable mixing angle $\alpha \sim \pi / 4$. Fig.~\ref{fig:bulkrelic} shows the characteristic parameter space where models with slepton mediators can satisfy the relic density through bino annihilation, with values of  $g_\mu - 2$ (described in Section~\ref{sec:dip}) consistent with experiment.

Models contained within the bulk region of the CMSSM can yield the observed DM relic density through bino annihilation mediated by \eg~scalar partners of the $\tau$~\cite{Pierce:2013rda}. Under the usual assumption of MFV in the CMSSM, only third generation scalars can have sufficient left-right mixing for the bino annihilation cross section to satisfy the relic density. However, if we relax the assumption of MFV then first and second generation scalars can have a large enough left-right mixing angle such that the relic density can be depleted to the observed value through bino annihilation. Referred to as the ``Incredible Bulk," such models incorporating scalar muon partners can yield interesting phenomenological signatures and potentially account for $g_\mu - 2$, as summarized in Sec.~\ref{sec:pheno}. 

\subsection{Co-annihilation} \label{sec:coann}

While the DM relic density can also be sufficiently depleted through bino annihilation in similar models with light flavor squark mediators, both limits from LHC on squark masses and on squark mediated DM scattering in direct detection experiments discussed in Sec.~\ref{sec:pheno} severely constrain such scenarios. On the other hand, for models with either squark or slepton mediators which are nearly degenerate in mass with the bino, the effects of co-annihilation can further expand the viable parameter space of charged mediator models. Since the characteristic mass scale of the spectra in such models can be closer to $m_{\s{B}} \simeq m_{\s{f}} \simeq \O(1) \, \TeV$ and the decay products of the mediators are more difficult to detect at the LHC, the next generation of direct detection experiments and future collider experiments currently under consideration may be the only probes sensitive to the full parameter space of charged mediator models.

In principle, the effects of co-annihilation must be calculated with a set of coupled Boltzmann equations which account for all interactions between the particle species that constitute the nearly mass degenerate spectrum and the thermal plasma in the early universe. However, since the relevant charged mediators in the spectrum are typically expected to decay to the bino soon after it decouples, we can write the set of coupled Boltzmann equations as one evolution equation 
\begin{equation}
    \frac{dn}{dt} = - 3 H n - \VEV{\sigma_{\rm eff} v} \left(n^2 - \left[n^{\rm eq} \right]^2 \right) \, ,
\end{equation}
where $n$ ($n^{\rm eq}$) is a sum of (equilibrium) number densities for all species in the nearly degenerate spectrum, and $H$ is the Hubble rate. The effective thermally averaged annihilation cross section is defined as
\begin{equation} \label{eqn:coannx}
   \VEV{\sigma_{\rm eff} v} = \sum_{i,j} \frac{n^{\rm eq}_i n^{\rm eq}_j}{\left[n^{\rm eq} \right]^2} \VEV{\sigma_{i j} v} \, ,
\end{equation}
where the contribution from the annihilation cross section of each initial state, $i,j=\{\s{B},\s{f}\}$, is weighted by the respective equilibrium number densities of the initial state species, $n_{i,j}$. Note that, similar to the case of bino annihilation, the DM relic density is satisfied for $\VEV{\sigma_{\rm eff} v} \sim 1 \, {\rm pb}$, again corresponding to a freeze-out temperature $T_f \sim m_{\s{B}}/25 $.

\begin{figure}
\begin{center}
\includegraphics[width=0.49\hsize]{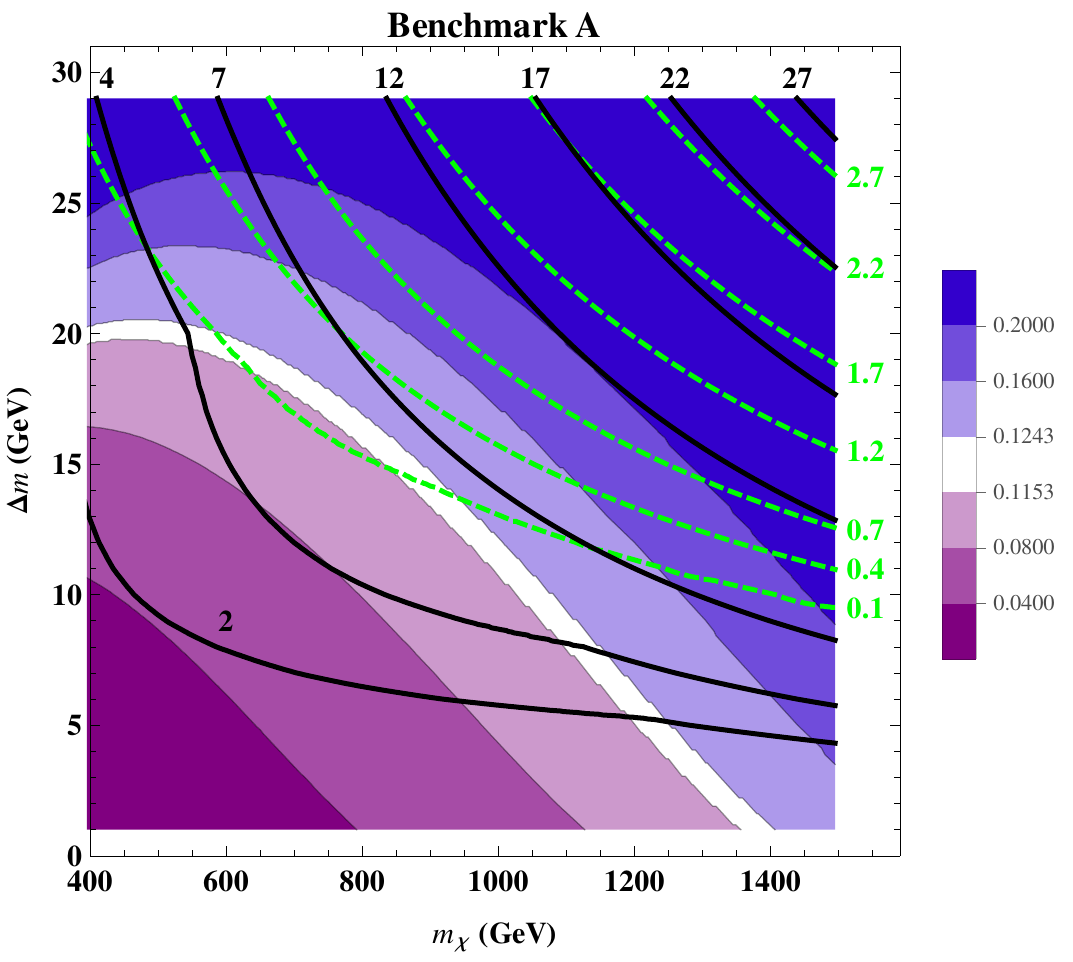}
\includegraphics[width=0.49\hsize]{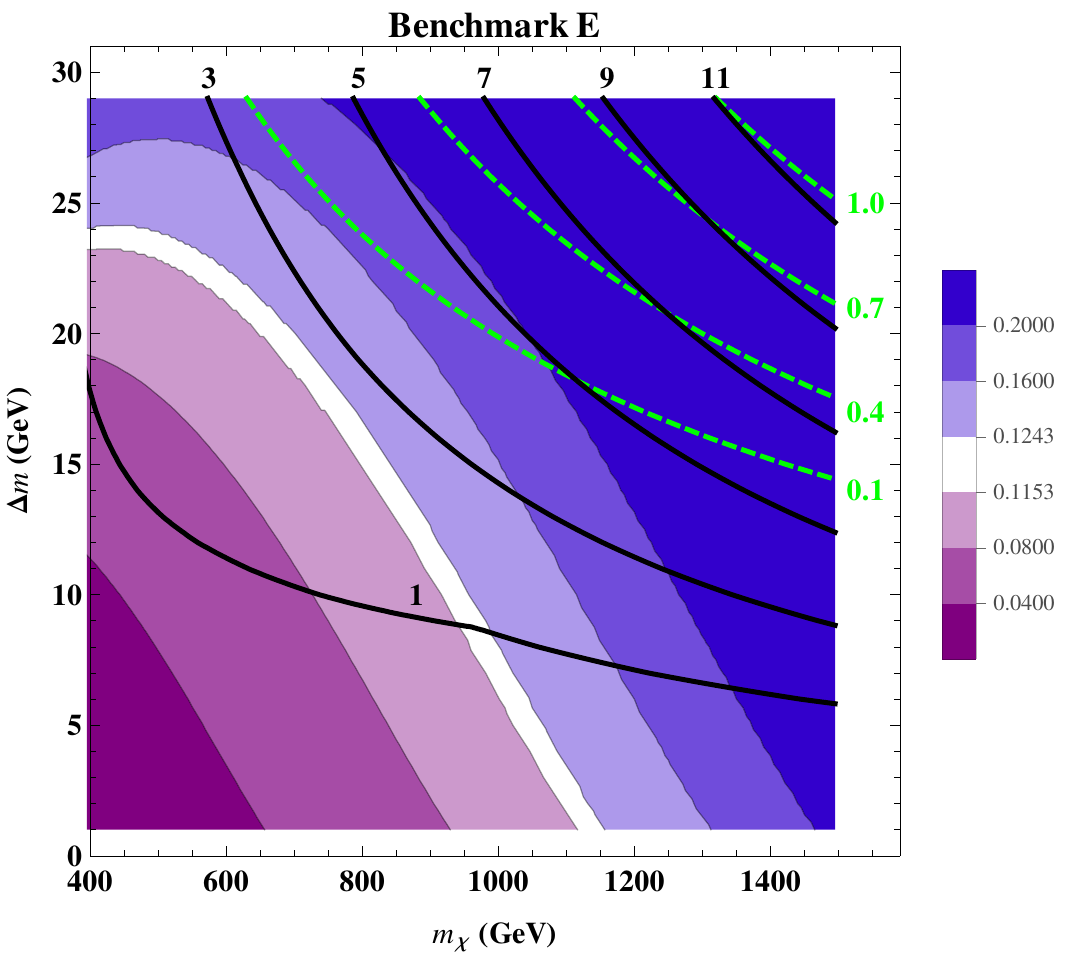}
\end{center}
\caption{Relic density contours for benchmark models with a single light flavor squark ($\s{u}_1$; left) and three mass degenerate light flavor squarks ($\s{u}_1,\s{d}_1,\s{s}_1$; right) as a function of bino mass $m_{\chi}$ and bino-squark mass splitting $\Delta m$. Black (green) lines show (projected) direct detection 90\% CL constraints  from XENON1T~\cite{XENON:2017vdw} (LZ~\cite{Mount:2017qzi}) for the values of the common squark left-right mixing angle $\alpha$ in units of $10^{-4}$, with all parameter space below the curves excluded. Appear in Fig.~4 of Ref.~\cite{Davidson:2017gxx}.}
\label{fig:AErelic}
\end{figure}

In order to briefly summarize the most relevant features of co-annihilation in charged mediator models, we focus on the simpler case of light flavor squarks nearly degenerate in mass with the bino. Due to the direct coupling of the bino to nucleons in such models, constraints from direct detection experiments described in Sec.~\ref{sec:pheno} restrict the chiral mixing angle to be small. In contrast, particularly for models which can satisfy $g_\mu - 2$, the chiral mixing of sleptons can be large and have a significant impact on the relic density calculation. Also, due to the strong gauge coupling of squarks to gluons, the process $\s{q}^*\s{q} \to g g$ typically provides the dominant contribution to $\VEV{\sigma_{\rm eff} v}$, which is independent of the chiral mixing angle, $\alpha$, or the flavor of the relevant squarks.

For a variety of light flavor squark combinations contributing to the effective annihilation cross section described by Eq.~\leqn{eqn:sqsqx} in Appendix~\ref{app:coann}, the DM relic density can be satisfied for $m_{\s{B}} \simeq \O(1) \, \TeV$ and $\Delta m = m_{\s{q}} - m_{\s{B}} \simeq \O(10) \, \GeV$. As shown in Fig.~\ref{fig:AErelic}, the Boltzmann suppression of the squark annihilation cross section is minimized for smaller $\Delta m$ and the squark mass must correspondingly increase to keep the relic density constant. In particular, squark annihilation can sufficiently deplete the relic density for $m_{\s{q}} \lesssim 1.4 \, \TeV$, well beyond the sensitivity of LHC for compressed spectra. The sensitivity of direct detection searches to these models is summarized in Sec.~\ref{sec:DD}. 

\begin{figure}
\begin{center}
\includegraphics[width=0.49\hsize]{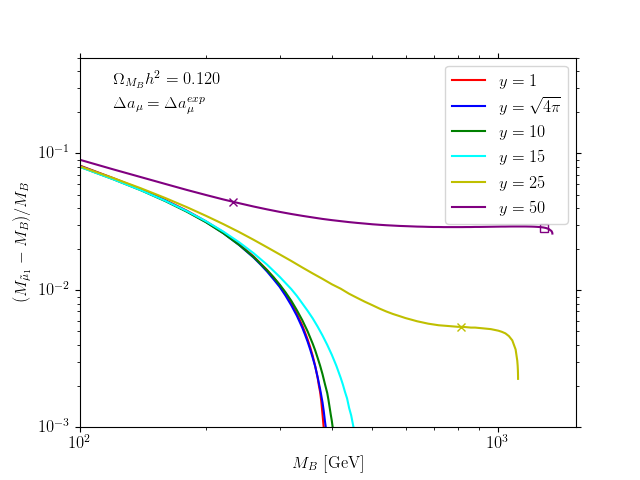}
\includegraphics[width=0.49\hsize]{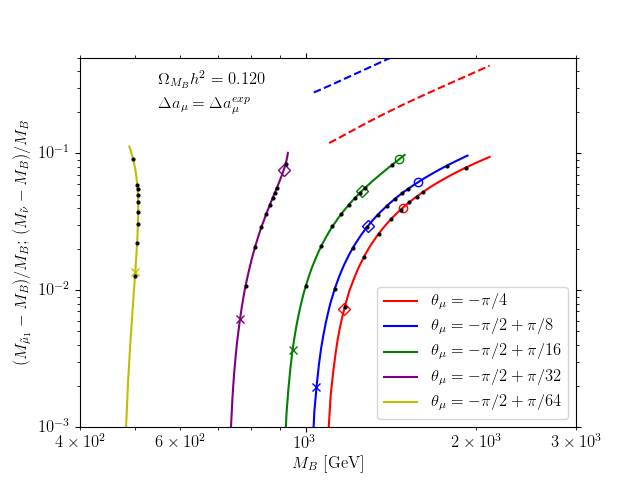}
\end{center}
\caption{Lines of fixed smuon mass splitting parameter $y=(M_{\s{\mu}_2}^2-M_{\s{\mu}_1}^2) | \sin ( 2 \theta_\mu ) | /(4M_W^2)$ (left) and left-right smuon mixing angle $\theta_\mu$ (right) for models which satisfy both the DM relic density and $g_\mu - 2$. In the right panel, the black dots along each line indicate increasing values of $y= 30, 40, 50, 60, 70, 80, 90, 100, 200$ as the relative bino-smuon mass splitting increases. For $y$ associated with a trilinear coupling between the Higgs boson and the smuons, the ``x" along each line indicates constraints from electroweak vacuum stability giving an upper limit on $M_{\s{B}}$ (left) and $y$ (right). Left and right panels appear in Figs.~2 and~3 of Ref.~\cite{Acuna:2021rbg}.
}
\label{fig:RBrelic}
\end{figure}

As mentioned above, slepton co-annihilation scenarios which can accommodate $g_\mu - 2$ measurements in addition to the relic density can be significantly more complicated. First, since sleptons are not charged under QCD, there typically is no single dominant contribution to the effective annihilation rate. Also, as described in Sec.~\ref{sec:dip}, accommodating $g_\mu - 2$ measurements in models with with smuons as heavy as $\sim 1 \, \TeV$ requires significant left-right smuon mixing and a large relative mass splitting between the smuons, characterised by the mass splitting parameter $y=(M_{\s{\mu}_2}^2-M_{\s{\mu}_1}^2) | \sin ( 2 \theta_\mu ) | /(4M_W^2)$. Unlike in squark co-annihilation models for which the mixing angle (here denoted $\theta_\mu$) is constrained to be small, cross sections for processes such as $\s{\mu}^*\s{\mu} \to W^+ W^-$, relevant for depleting the relic density in slepton co-annihilation scenarios, can exhibit peculiar scaling behaviors in the non-relativistic limit as the slepton mass scale increases. As shown for models with a large mass splitting parameter $y$ in Fig.~\ref{fig:RBrelic}, the Boltzmann suppression necessary to yield the correct relic density can remain constant or even increase with $M_{\s{B}} \simeq M_{\s{\mu}_1}$. A detailed investigation of perturbative unitarity and electroweak vacuum stability in such models can be found in Ref.~\cite{Acuna:2021rbg}.

\section{Phenomenology} \label{sec:pheno}
 
In this section, we summarize the most interesting phenomenological features of charged mediator models. Similar to Appendix~\ref{app:relic} for Sec.~\ref{sec:relic}, we leave several more detailed equations and explanations to Appendix~\ref{app:pheno}. For complete and self-contained discussions of dipole moments, collider searches, direct detection and indirect detection in the context of charged mediator models, please see previous work in respective Refs.~\cite{Fukushima:2013efa,Fukushima:2014yia},~\cite{Dutta:2014jda,Dutta:2017nqv,Baum:2020gjj},~\cite{Kelso:2014qja,Sandick:2016zut,Davidson:2017gxx,Acuna:2021rbg} and~\cite{Kumar:2016cum,Sandick:2016zeg}.

\subsection{Dipole moment constraints} \label{sec:dip}

\begin{figure}
\begin{center}
\includegraphics[width=0.25\hsize]{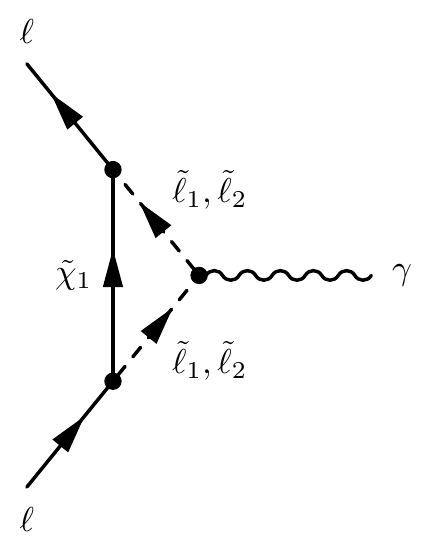}
\end{center}
\caption{Feynman diagram for dipole moment contributions in charged mediator models. Appears as Fig.~2 of Ref.~\cite{Fukushima:2014yia}.
}
\label{fig:dipolediag}
\end{figure}

If a Standard Model fermion ($f$) couples to dark matter through interactions with charged mediators, then these interactions yield a one-loop contribution to the magnetic and electric dipole moments  of the SM fermion, with the dark matter and mediators running in the loop (see Fig.~\ref{fig:dipolediag}). Interestingly, we find as a general principle that the corrections to the dipole moments can be related the cross section for $s$-wave bino annihilation, $\s{B} \s{B} \to \bar f f$. Comparing the contribution to the annihilation cross section in Eq.~\leqn{eq:c0} of Appendix~\ref{app:bulk} to the dipole moments in Eq.~\leqn{eq:DMs} of Appendix~\ref{app:dip}, each expression contains both the left-handed and right-handed couplings ($\lambda_{L,R}$), as well as the mixing angle.

Majorana fermion dark matter $s$-wave annihilation can only occur from a $J^{PC} = 0^{-+}$ initial state (see, for example,~\cite{Kumar:2013iva}). The correction to the magnetic dipole moment is related to the $CP$-conserving matrix element for annihilation to the $J^{PC} = 0^{-+}$ final state, while the correction to the electric dipole moment is related to the $CP$-violating matrix element for annihilation to the $J^{PC} = 0^{++}$ final state~\cite{Fukushima:2013efa,Fukushima:2014yia}. In particular, in the limit where the mass splitting between the dark matter and mediator is large, we find~\cite{Fukushima:2014yia}
\begin{equation}
c_0 \sim 3.9 \times 10^{11}~{\rm pb} \left[ (\Delta a_f)^2 + \left(\frac{2m_f d_f}{|e|} \right)^2 \right] 
\left( \frac{m_f}{\GeV} \right)^{-2} .
\label{eq:DipoleConstraints}
\end{equation}
Thus, experimental measurements of the electric and magnetic dipole moments constrain the $s$-wave bino annihilation cross section in charged mediator models. Note, if the lighter scalar mass eigenstate is taken to be nearly degenerate in mass with the bino while keeping the heavier mass eigenstate decoupled, then the $s$-wave annihilation cross section is maximized and the bound in Eq.~\leqn{eq:DipoleConstraints} is weakened by about a factor of 2.

\begin{figure}
\begin{center}
\includegraphics[width=0.35\hsize]{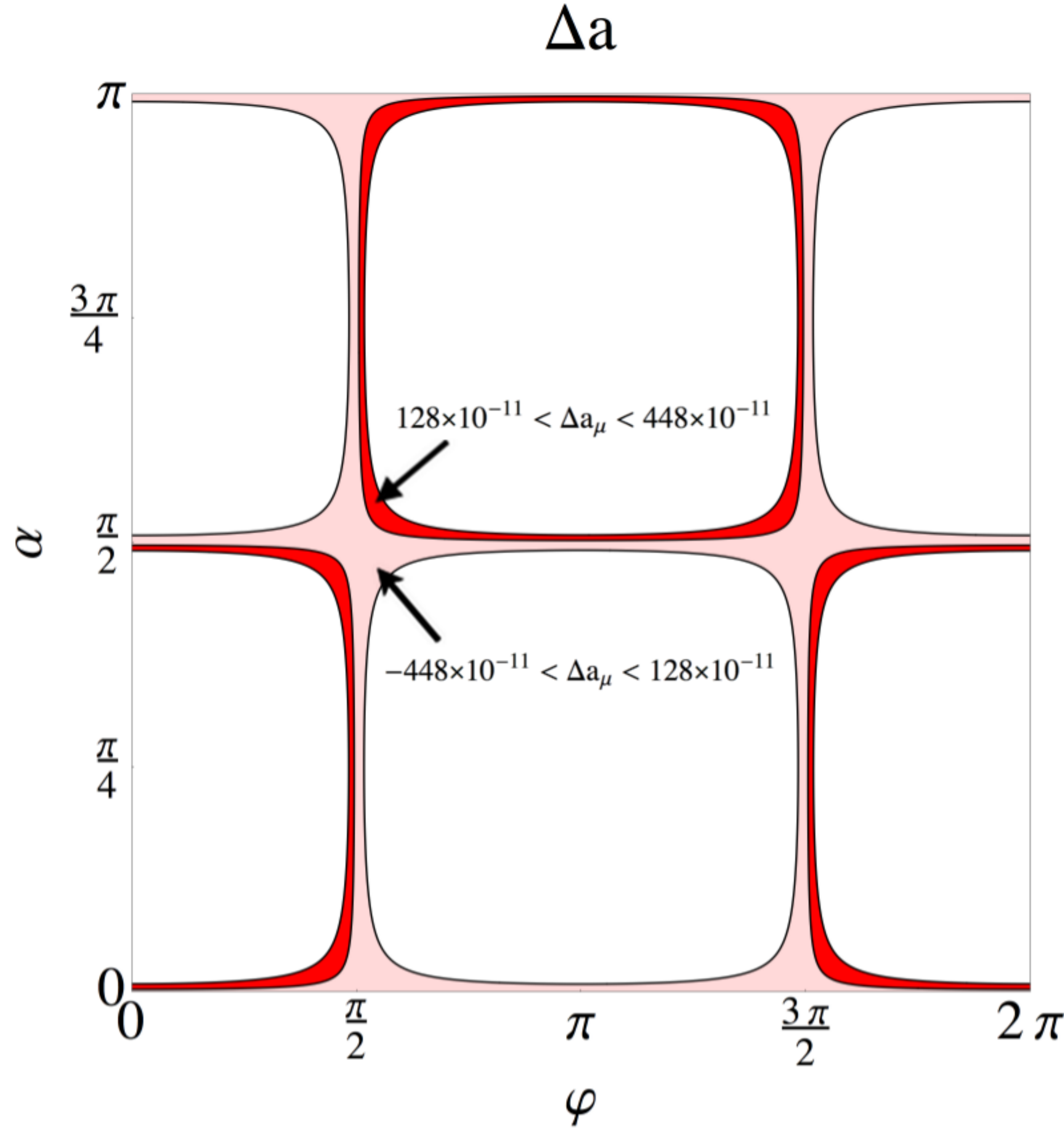}
\includegraphics[width=0.49\hsize]{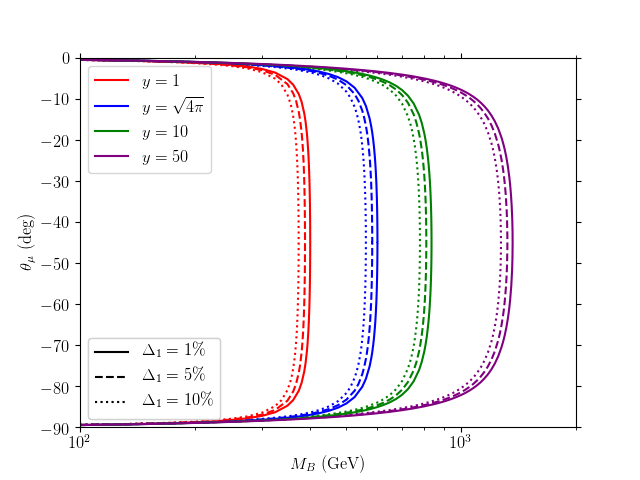}
\end{center}
\caption{{\it Left} Dependence of $\Delta a_\mu$ on left-right mixing angle, $\alpha$, and $CP$-violating phase, $\phi$, for models with a $m_{\s{B}} = 100 \, \GeV$ bino coupled to muons through scalar muon partners with masses $m_{\s{\mu}_1} = 120 \, \GeV$ and $m_{\s{\mu}_2}  = 300 \, \GeV$. Appears in Fig.~4 of Ref.~\cite{Fukushima:2014yia}. {\it Right} Setting the $CP$-violating phase $\phi = 0$ and focusing on models with a small relative mass splitting between the bino and lightest smuon $\Delta_1 = (M_{\s{\mu}_1} - M_{\s{B}})/M_{\s{B}}$, the parameter space consistent with $\Delta a_\mu$ extends out to larger mass scales as the left-right mixing angle, $\theta_\mu$, tends towards maximal and the smuon mass splitting parameter, $y$, increases. Appears in Fig.~1 of Ref.~\cite{Acuna:2021rbg}.
}
\label{fig:amu}
\end{figure}

As a practical matter, current measurements of quark and $\tau$ dipole moments are too weak to provide useful constraints on $s$-wave dark matter annihilation~\cite{Fukushima:2014yia,Kelso:2014qja}. On the other hand, corrections to the dipole moments of the electron are so tight as to rule out entirely scenarios in which $\s{B} \s{B} \to e^+ e^-$ proceeds from an $s$-wave state. If the bino couples to the muon, we find that measurements of the muon anomalous magnetic moment tightly constrain the $CP$-conserving annihilation matrix element. Instead, $s$-wave annihilation to muons must proceed through an interaction which is almost entirely $CP$-violating. The left-right mixing angles and $CP$-violating phases most relevant for models with scalar muon partners accounting for $g_\mu - 2$ are shown in the left panel of Fig.~\ref{fig:amu}.

Note that the cross section for $p$-wave annihilation can similarly be related to corrections to the vertex function which are higher order in momentum~\cite{Kumar:2016gxq}, namely, to the charge radius. Focusing on the case $f=\mu$, we see that corrections to the charge radius can be probed by searches for lepton non-universality.  Current bounds from LEP~\cite{L3:2000bql} do not lead to constraints on the $p$-wave annihilation cross section strong enough to be of interest, but future high-energy lepton colliders, such as ILC or CLIC, could provide much greater sensitivity.

Alternative to models with scalar muon partners which satisfy the relic density through bino annihilation, $g_\mu - 2$ measurements can also be accounted for in models with $M_{\s{\mu}_1}$ sufficiently close to $M_{\s{B}}$ such that the effects of co-annihilation are relevant. With a less direct relationship to relic density than implied by Eq.~\leqn{eq:DipoleConstraints}, such models do not require a $CP$-violating phase to reconcile the relic density with $g_\mu - 2$, so we fix $\phi = 0$. In the limit where $M_{\s{B}} \simeq M_{\s{\mu}_1} \gg M_W$ and assuming $M_{\s{\mu}_2}^2 - M_{\s{\mu}_1}^2 \simeq M_W^2$, we can approximate~\cite{Acuna:2021rbg}
\begin{equation}
  \frac{\Delta a_\mu}{25.1 \times 10^{-10}} \simeq \left( \frac{y}{10} \right) \left(\frac{1 \TeV}{M_{\s{B}}} \right)^3 \left(\frac{1+0.24(\Delta_1/0.1)}{1.24} \right) \, ,
\end{equation}
with the relative bino-smuon mass splitting typically $\Delta_1 = (M_{\s{\mu}_1} - M_{\s{B}})/M_{\s{B}} \lesssim 5 - 10 \%$ to satisfy the relic density. In the right panel of Fig.~\ref{fig:amu}, we can see that $\Delta a_\mu$ can indeed match the observed value for $M_{\s{B}} \simeq M_{\s{\mu}_1} \sim 1 \, \TeV$ with $y \gg 1$ and significant left-right smuon mixing. As $M_{\s{B}}$ decreases for fixed $y$, $g_\mu - 2$ can only be satisfied for light smuon mass eignestates with close to purely left- or right-handed chirality. We refer to these as the left- or right-handed ``branches" of the slepton co-annihilation parameter space, primarily focusing on the latter in this white paper.

\subsection{Collider searches} \label{sec:coll}

In Section~\ref{sec:bulk} we show that in the Incredible Bulk region the DM relic density can be satisfied if $m_{\s{B}} \simeq \mathcal{O}(100)$ GeV and $\Delta m=m_{\s{f}_1}-m_{\s{B}} \lesssim 60$ GeV. If $\s{f}_1=\s{\mu}_1$ then $g_{\mu}-2$ can also be satisfied as can be seen from Fig.~\ref{fig:bulkrelic}. Interestingly, from Fig.~\ref{fig:AErelic} and Fig.~\ref{fig:RBrelic} we can infer that in the co-annihilation regime the DM abundance can be similarly satisfied with $\Delta m \lesssim 30$ GeV (also $g_{\mu}-2$ for smuon mediators) for $m_{\s{B}} \lesssim 1.4$ TeV. 
 
These compressed spectra scenarios favored by DM relic density considerations are notoriously difficult to probe at the LHC. This is because, in compressed SUSY spectra, the leptons and jets produced from sfermion NLSP decays will be soft and will not pass the respective $p_T$ thresholds. The missing energy generated in the sfermion pair system will also be small and will not provide us any handle to suppress the SM BGs. So, the impressive bounds that CMS and ATLAS have imposed on slepton and squark pair production at the LHC will not apply to compressed scenarios as these analyses fail for $\Delta m \lesssim 60$ GeV~\cite{ATLAS:2019lff,CMS:2020bfa}. Hence, one needs to boost a compressed spectra system by means of one (monojet) or two (VBF) energetic jets. The boost will not only enhance the lepton and jet transverse momenta but also, to balance the momentum of the jet(s), a larger opening angle will be created between the LSPs resulting in higher $E_T^{\text{miss}}$.

Before going into the details of monojet and VBF searches let us briefly discuss the prospect of sfermions being long-lived at the LHC. If the sleptons are stable on the LHC-detector timescale they will be long-lived charged particles and will behave like muons. However, they can be differentiated from muons from time-of-flight measurements in the muon spectrometer and energy loss in the inner detectors. For example, using these techniques ATLAS excluded $m_{\s{\ell}}\sim 377-335$ GeV in models where the decays of heavier sleptons produced at $\sqrt{s}=8$ TeV yield a long-lived stau NLSP, assuming mass-splittings $m_{\s{\ell}}-m_{\s{\tau_1}}=2.7 - 93$ GeV~\cite{ATLAS:2014fka}. Using only the time-of-flight information Ref.~\cite{Feng:2015wqa} projects that the LHC will be able to exclude $m_{\s{\ell}}\sim 1.2$ TeV at $\sqrt{s}=14$ TeV with 3 ab$^{-1}$ of integrated luminosity. With the same luminosity they also estimate that a 100 TeV $pp$ collider can exclude $m_{\s{\ell}}\sim 4$ TeV. 

On the other hand, if sleptons decay after traveling some distance within the LHC detectors, they will give rise to displaced lepton signatures. Here, one can not trace the final state leptons back to the collision point. In a recent displaced lepton analysis~\cite{ATLAS:2020wjh}, ATLAS excludes $m_{\s{e}} \, (m_{\s{\mu}}) \sim 720 \, (680)$ GeV using 139 fb$^{-1}$ luminosity for lifetimes of 0.1 ns. As far as the long-lived searches for the first and second-generation squarks are concerned, CMS has recently performed a disappearing track search~\cite{CMS:2019ybf} for them assuming all eight light flavor squark states are degenerate. However, they assume the LSP to be wino-like while our LSP is bino-like. So, we do not quote the limits from that study in this white paper.
 
Let us now come back to monojet and VBF searches for the first two generation sfermions at the LHC. First, we discuss sleptons. Refs.~\cite{Han:2014aea,Barr:2015eva} tried to understand compressed slepton spectra using monojet+soft di-lepton+$E_T^{\text{miss}}$ final state, while Ref.~\cite{Dutta:2014jda} did the same with di-jet+$E_T^{\text{miss}}$+ one or two soft leptons using the VBF topology. The LHC experiments adopted some of the ideas proposed in the above papers to search for compressed slepton spectra. Using 139 fb$^{-1}$ of accumulated luminosity at $\sqrt{s}=13$ TeV, ATLAS excludes slepton masses up to 251 GeV with $\Delta m \sim 10$ GeV, assuming all four first two generation sleptons are degenerate~\cite{ATLAS:2019lng}. However, at the extreme ends of mass-splitting probed by ATLAS, namely with $\Delta m \sim 550$ MeV and 30 GeV, the slepton mass bounds relax to existing LEP limits of 73 and 100 GeV, respectively. In contrast, CMS studied compressed slepton spectra with VBF search strategy only~\cite{CMS:2019san}. Due to the lower production cross-section of VBF compared to monojet, VBF limits are much weaker with the current data available and do not improve on LEP bounds~\cite{ATLAS:2019lng,CMS:2019san}.

However, none of the above search strategies work for $\Delta m \sim 30-60$ GeV. The main BG for the compressed slepton searches are $t \bar{t}, \, V$ and $VV$ + jets. The above analyses critically depend on an upper bound on $p_T$ of leptons or an upper bound on related kinematic variables $m_{\ell \ell}$ and $M_{T_2}$. These variables lose their efficacy for $\Delta m > 30$ GeV since the lepton $p_T$ distributions for the SM BGs, coming from $W, \, Z$ decays, peak around 40 GeV, and it becomes difficult to differentiate between signal and BG distributions. 

In Ref.~\cite{Dutta:2017nqv} we propose additional measures to suppress the SM backgrounds further to probe $\Delta m \sim 30-60\gev$ for smuons. Since the signal rate for monojet is higher than VBF we focus on monojet+di-lepton+$E_T^{\text{miss}}$ final state. In addition to well known discriminators of $p_{T_j},\, E_T^{\text{miss}}$ and $m_{\tau \tau}$ for this channel, we use the variable $\cos \theta_{\ell_1,\ell_2} \equiv \tanh{\Delta \eta_{\ell_1,\ell_2}}$. This variable exploits the spin of the mother particle to distinguish them, and is helpful to suppress $WW$ and $t \bar{t}$ backgrounds by a factor of two. Next, we use different strategies for $\Delta m < 50$ GeV and $\Delta m \sim 50 -60$ GeV. In both cases the main challenge is to beat $WW$ + jets. 

\begin{figure}
\begin{center}
\includegraphics[width=0.3\hsize]{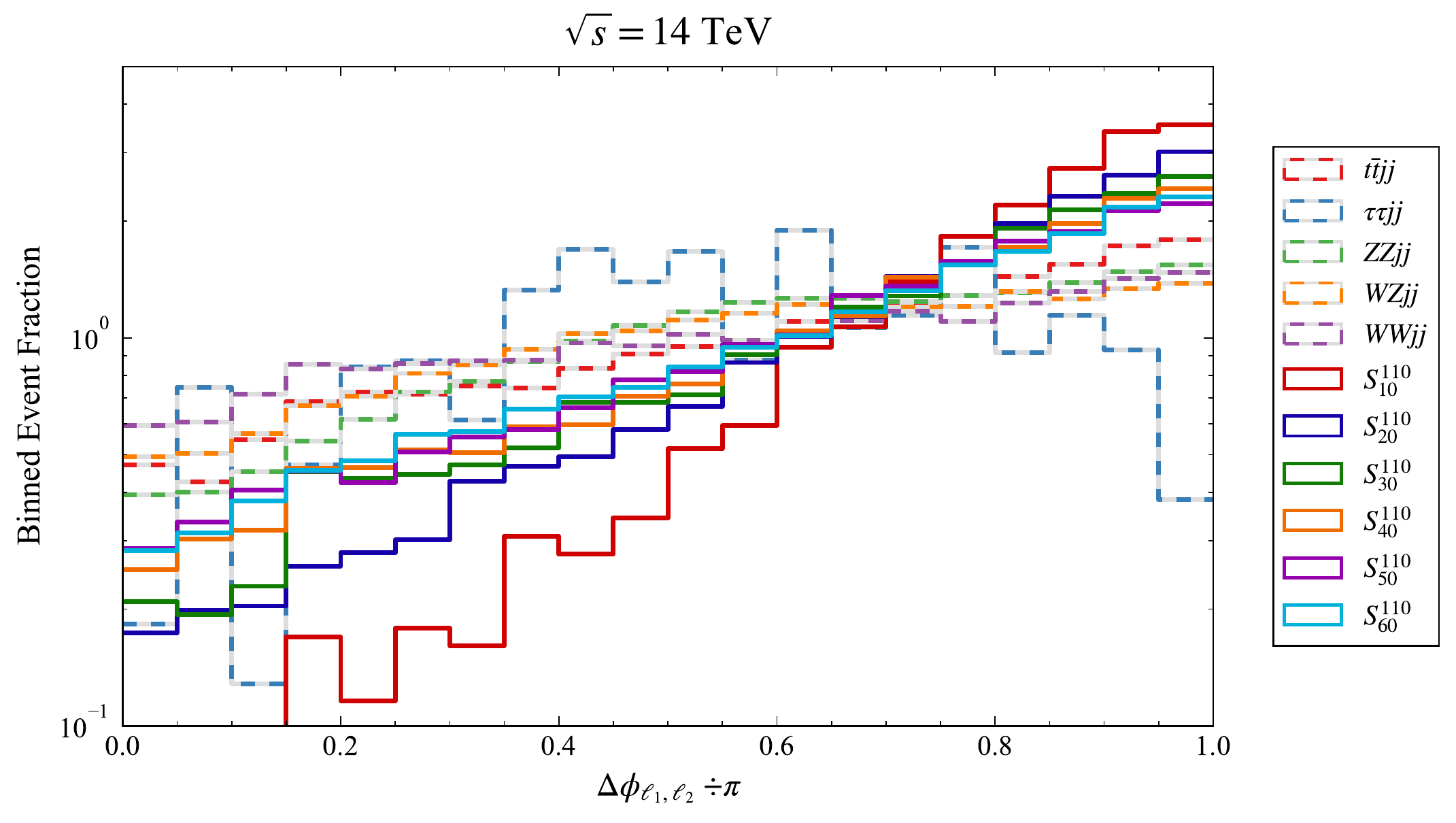}
\includegraphics[width=0.3\hsize]{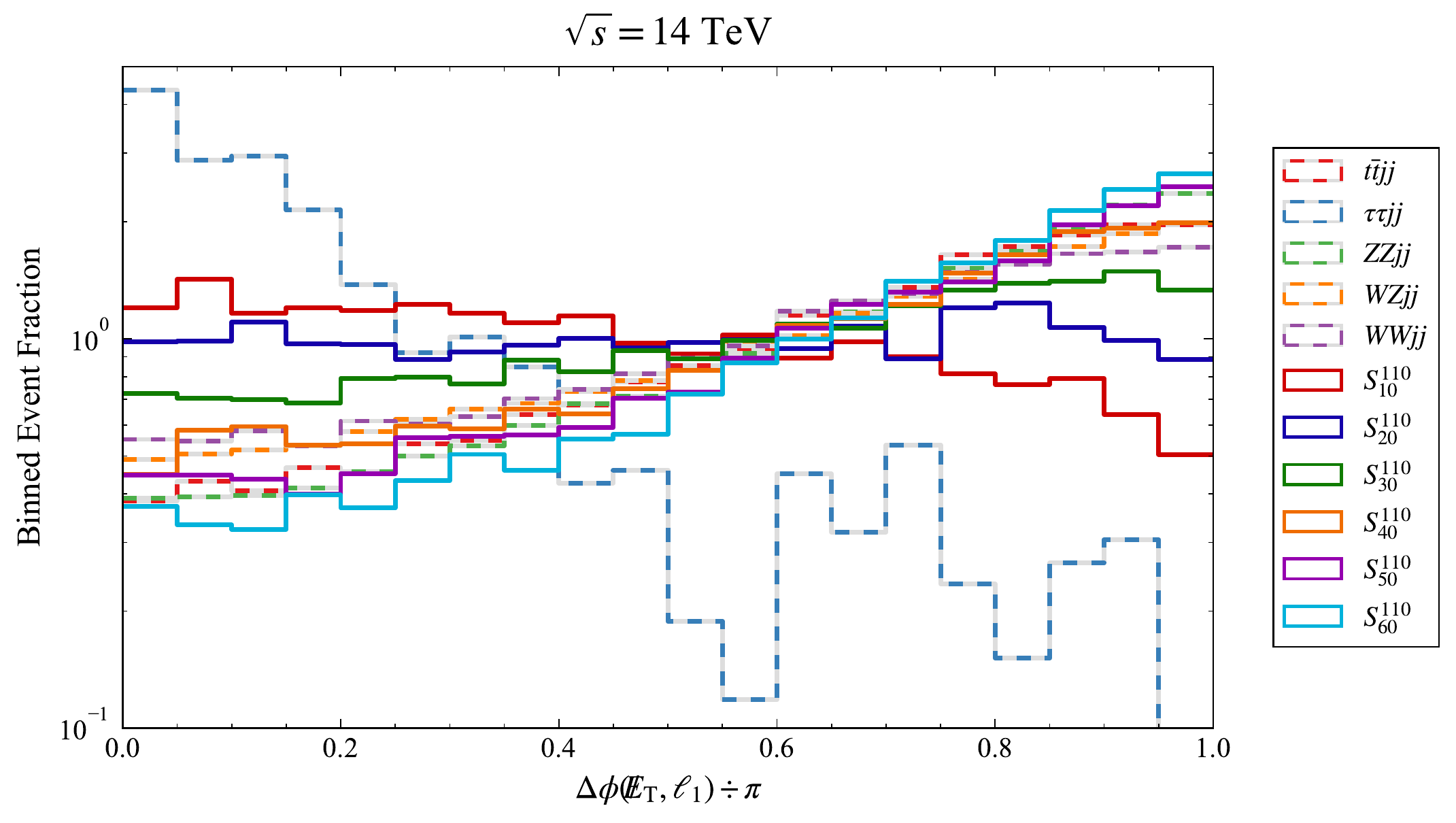}
\includegraphics[width=0.3\hsize]{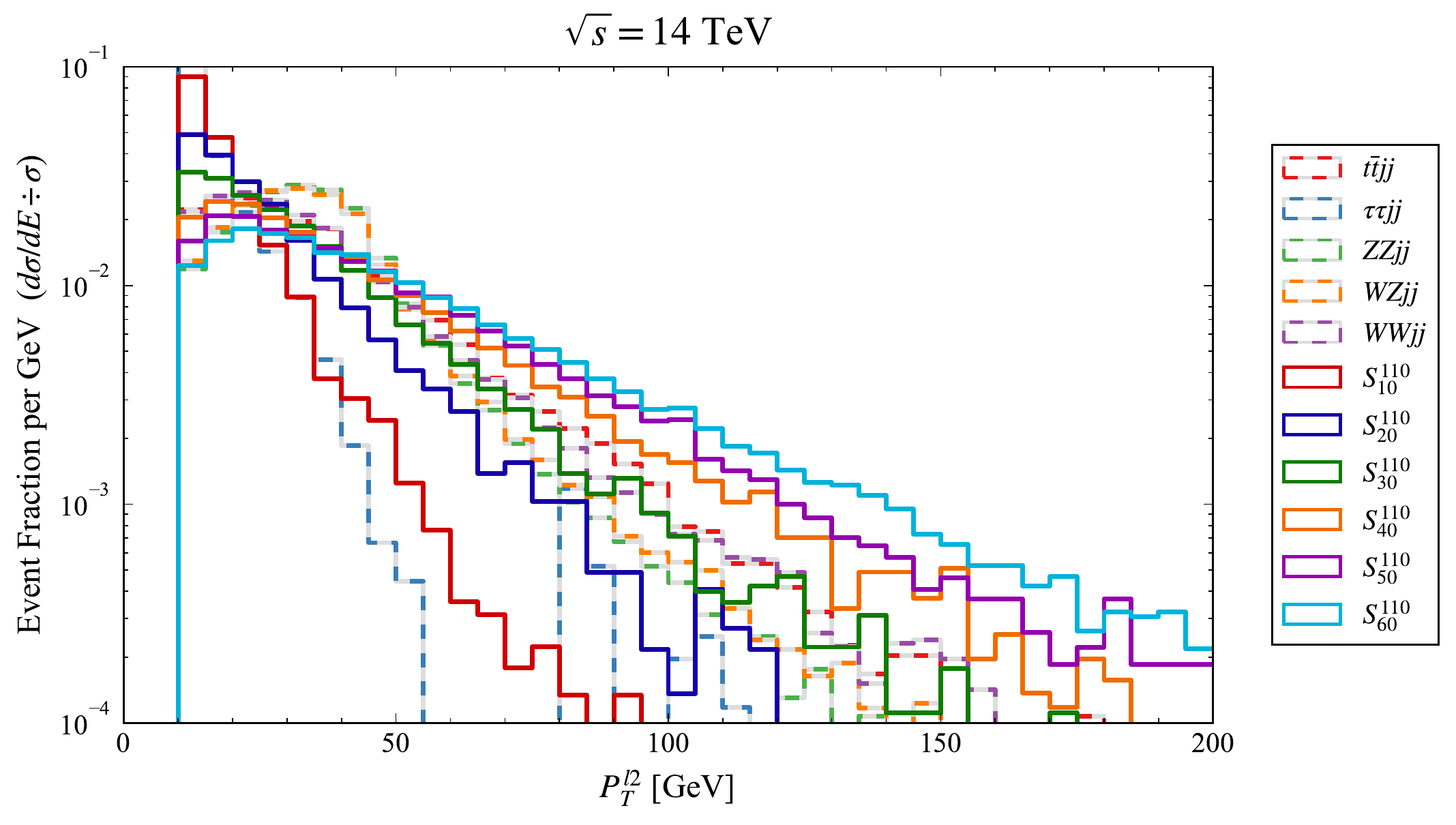}
\end{center}
\caption{Signal and background event shapes are compared for the azimuthal angular separation between the visible di-lepton pair (left), the azimuthal angular separation between the missing transverse energy and the leading lepton $\Delta \phi({\ell_{1}, E_T^{\text{miss}}})$ (center), and the sub-leading lepton transverse momentum $p_T^{\ell_2}$ (right). Appear in Figs.~3 and~4 of Ref.~\cite{Dutta:2017nqv}.
}
\label{fig:LHCkin}
\end{figure}

For intermediate mass splittings ($20 - 40$ GeV), we make use of the cuts of $\Delta \phi({\ell_1,\ell_2})/\pi > 0.5$ and $\Delta \phi({\ell_{1,2}, E_T^{\text{miss}}})/\pi <0.6$ to reduce $W,Z$ backgrounds further. If the mother particles ($\s{\mu}, W$ here) are produced with reasonable longitudinal boost, the daughter particles will be collimated with the mother, and the leptons will tend to be anti-collimated with each other. However, the transverse boost generated in the system by the emission of an energetic jet will smear these angular correlations. Relative to the production of less massive $W$-bosons, the angular correlations associated with smuon production are smeared out less. The $\Delta \phi({\ell_{1,2}, E_T^{\text{miss}}})$ cut is helpful for additional optimization of the signal by diminishing $t \bar{t}$ + jets, since a significant portion of $E_T^{\text{miss}}$ in this BG arises from mismeasured or missed jets. The $\Delta \phi({\ell_1,\ell_2})$ and $\Delta \phi({\ell_{1}, E_T^{\text{miss}}})$ distributions for the signal with $m_{\s{B}}=100$ GeV and different mass splittings, along with the SM BGs, are presented in the left and center panels of Fig.~\ref{fig:LHCkin}.

For larger mass gaps (50-60 GeV) some of our previous considerations do not apply, since with increasing mass splitting the di-lepton + $E_T^{\text{miss}}$ system becomes less collimated and harder than SM BGs. We found that a good significance can be obtained in this region by applying $\Delta \phi({\ell_{1,2}, E_T^{\text{miss}}})/\pi > 0.25$ and $p^{\ell_2}_T > 40$ GeV. In the right panel of Fig.~\ref{fig:LHCkin} we show the $p^{\ell_2}_T$ distributions for the signal with different mass gaps and BGs. To summarize, using the above strategies the LHC will be able to exclude $m_{\s{\mu}} \approx 200$ GeV at $1.5-2.3 \sigma$ level for $\Delta m \sim 30-60$ GeV with 1000 fb$^{-1}$ of integrated luminosity. One can further improve  LHC sensitivity by using machine learning and Bayesian optimization techniques~\cite{Alves:2017ued}.  

For the squark mediator cases, squark NLSP decays lead to soft jets. Although the LHC experiments were remarkably successful in lowering $p_T$ thresholds for light leptons, it is extremely difficult to lower the same for jets in the gluon rich environment of the LHC. However, the production cross-sections of QCD-charged squarks are much higher than for electroweak slepton particles. Consequently, one can efficiently probe compressed squark spectra using monojet and missing energy final states, which suffer from tiny signal to BG ratio for sleptons. The ATLAS collaboration excludes $m_{\s{q}} \sim 900 - 750$ GeV for $\Delta m \sim 5-50$ GeV utilizing this channel~\cite{ATLAS:2021kxv}. 

Given the difficulty of probing $\sim \TeV$ sleptons relevant for co-annihilation scenarios at the LHC (assuming the sleptons are not  long-lived on detector timescales), it is interesting to consider the sensitivity of a future lepton collider. The reconstruction of the soft leptons and smaller missing energy characteristic of signals associated with compressed spectra is considerably more straight forward at lepton colliders. In addition the lack of pileup and underlying event associated with final states at hadron colliders, lepton colliders are also planned to operate without triggers and, particularly for linear $e^+ e^-$ colliders such as the ILC, beam polarization can be optimised to suppress challenging backgrounds such as $WW$.

\begin{figure}[t]
\begin{center}
\includegraphics[width=0.49\hsize]{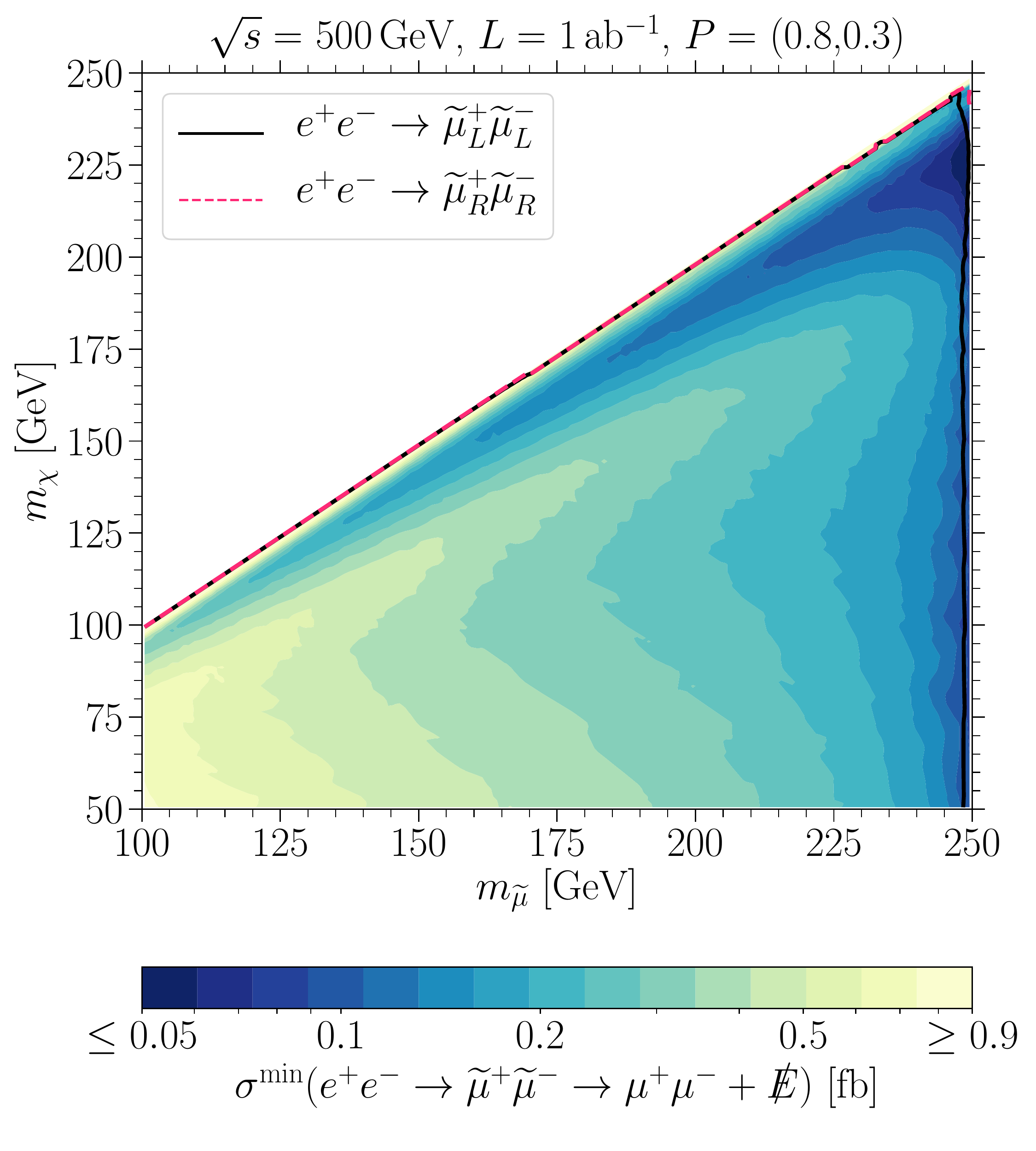}
\includegraphics[width=0.49\hsize]{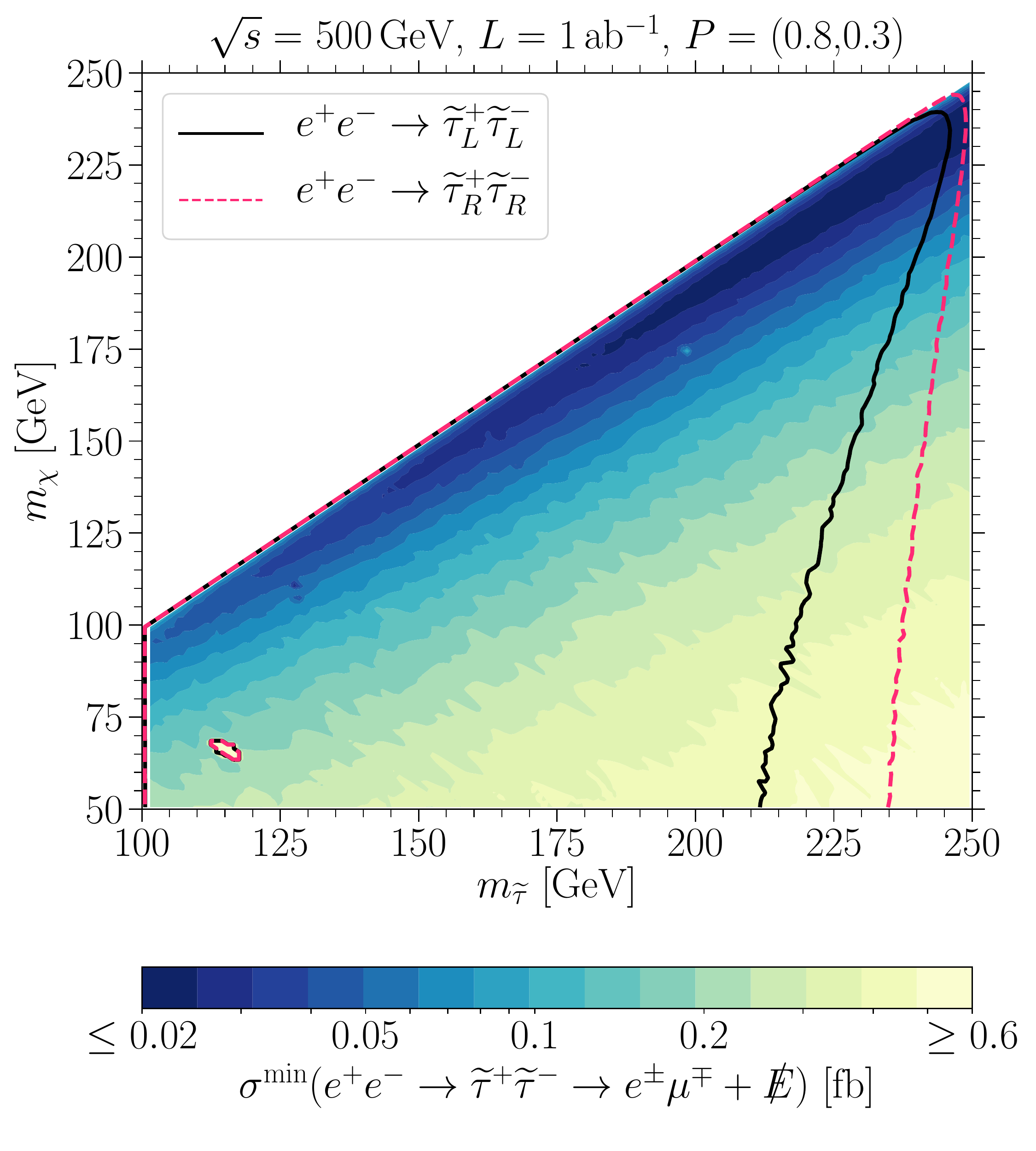}
\end{center}
\caption{Projected sensitivity for the smallest cross sections $\sigma_{\rm min}$ detectable for smuon (left) and stau (right) signals at a future $\sqrt{s} = 500 \, \GeV$ $e^+ e^-$ collider with integrated luminosity of $L = 1 \, {\rm ab}^{-1}$ and mostly right-handed beam polarization as a function of slepton and bino mass. For reference, the solid lines show the sensitivity for the production of purely left (solid black) and right (dashed red) chiral sleptons which decay with $100 \%$ branching fraction to the corresponding lepton and bino. Appears as Fig.~6 of Ref.~\cite{Baum:2020gjj}.}
\label{fig:ilc}
\end{figure}

In Ref.~\cite{Baum:2020gjj}, we investigate the potential sensitivity of searches for smuon and stau production at $e^+ e^-$ colliders. For smuon production we consider a final state with opposite signed muons and missing energy, while for stau searches we demonstrate that the most promising final state contains opposite sign leptons of mixed flavor and missing energy. Many of the kinematic features used to distinguish the final states associated with slepton production from the most relevant BGs are more easily recognizable at $e^+ e^-$ colliders since the CM frame of $e^+ e^-$ collisions is coincident with the detector frame. For example, the BG processes with the largest cross sections in smuon searches with a CM energy of $\sqrt{s} = 500 \, \GeV$ are $(e^+ e^- / \gamma \gamma \to \mu^+ \mu^-)$, with the $\gamma \gamma$ initial state processes induced by the photon component of the $e^+ e^-$ beams. Since these processes do not feature missing transverse energy, this BG can be suppressed by requiring $E^{\text{miss}} > 5 \, \GeV$, $\Delta \phi({\ell^+,\ell^-})/\pi < 0.8$ and $|\cos \theta (E^{\text{miss}}) | < 0.9$ (polar angle of $E^{\text{miss}}$ with respect to the beam axis).

The most significant BGs remaining after these initial cuts include processes with neutrinos in the final state, including $e^+ e^- \to WW$. Since the final state of the signal includes a pair of massive binos rather than neutrinos, the invariant mass of the associated lepton pair is considerably smaller than for $WW$ production and we impose a cut of $m_{\ell \ell} < 300 \, \GeV$ for $\sqrt{s} = 500 \, \GeV$. Also, since one of the largest channels for $WW$ production is $t$-channel neutrino exchange, the BG can be further suppressed by requiring $|\cos \theta (\ell_1) | < 0.8$. A full description of the optimal cut flow, including a final ``window" cut on the leading lepton momentum, for both smuon and stau searches at future $e^+ e^-$ colliders with a variety of CM energies, beam polarizations and integrated luminosities can be found in Ref.~\cite{Baum:2020gjj}.

As an illustrative example of the projected sensitivity of a $\sqrt{s} = 500 \, \GeV$ $e^+ e^-$ collider with a mostly right-handed beam polarization and integrated luminosity of $1 \, {\rm ab}^{-1}$, we show the contours of the minimum detectable smuon and stau signal cross sections in Fig.~\ref{fig:ilc}. The lines indicate the regions of the bino-slepton mass plane to which such a collider would be sensitive assuming the branching fraction for slepton decay to the corresponding lepton and bino is $100 \%$. We see that smuon searches at a collider similar to the proposed ILC could essentially cover all of the viable parameter space for charged mediator models up to $m_{\s{\mu}} \lesssim \sqrt{s} / 2$. The sensitivity of stau searches do not improve as $m_{\s{\tau}} \to \sqrt{s} / 2$ due to both the branching fractions and kinematics associated with stau decay. However, particularly for $e^+ e^-$ colliders such as CLIC with proposed CM energies as high as $\sqrt{s} \simeq 3 \, \TeV$, we see that a future lepton collider could be an ideal probe for charged mediator models.

\subsection{Direct detection} \label{sec:DD}

If dark matter couples to quarks through charged mediators, then dark matter-nucleon scattering can proceed through exchange of a charged mediator in the $s/u$-channel. The matrix element can be expressed in terms of a variety of dimension-6 effective operators whose coefficients are determined by the parameters of the model (see Appendix~\ref{app:DD}). The dimension-6 operator which mediates spin-independent (SI) velocity-independent scattering has a coefficient which is proportional to the mixing angle. As a result, in scenarios respecting MFV, the SI scattering cross section is suppressed, and is dominated by coupling to the heaviest partons.  This implies that, in scenarios respecting MFV, the SI scattering cross section respects isospin as well.

But in the broader scenario we consider, in which flavor violation can be non-minimal, larger SI scattering cross sections are allowed.  Moreover, since the mediators can couple to any quark and still have a large mixing angle, there can be significant SI scattering through interactions with first-generation quarks, naturally leading to isospin-violation~\cite{Chang:2010yk,Feng:2011vu,Feng:2013vod}.

In the limit of negligible scalar mixing, the dominant velocity-independent term in the matrix element generated by dimension-6 operators is spin-dependent. However, there is a twist-2 dimension-8 operator which can mediate velocity-independent SI scattering even in the limit of small left-right mixing. Although this operator is suppressed by higher powers of the mediator mass, the coherent enhancement of the SI scattering cross section can allow it to dominate the scattering cross section for small mixing.

\begin{figure}[t]
\begin{center}
\includegraphics[width=0.49\hsize]{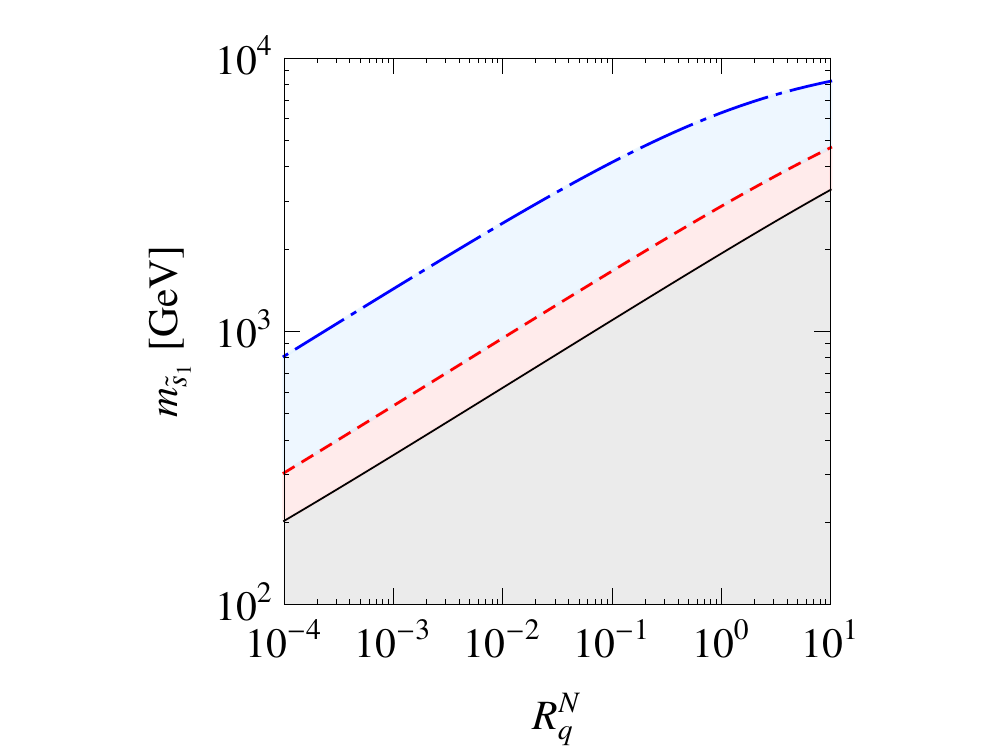}
\includegraphics[width=0.49\hsize]{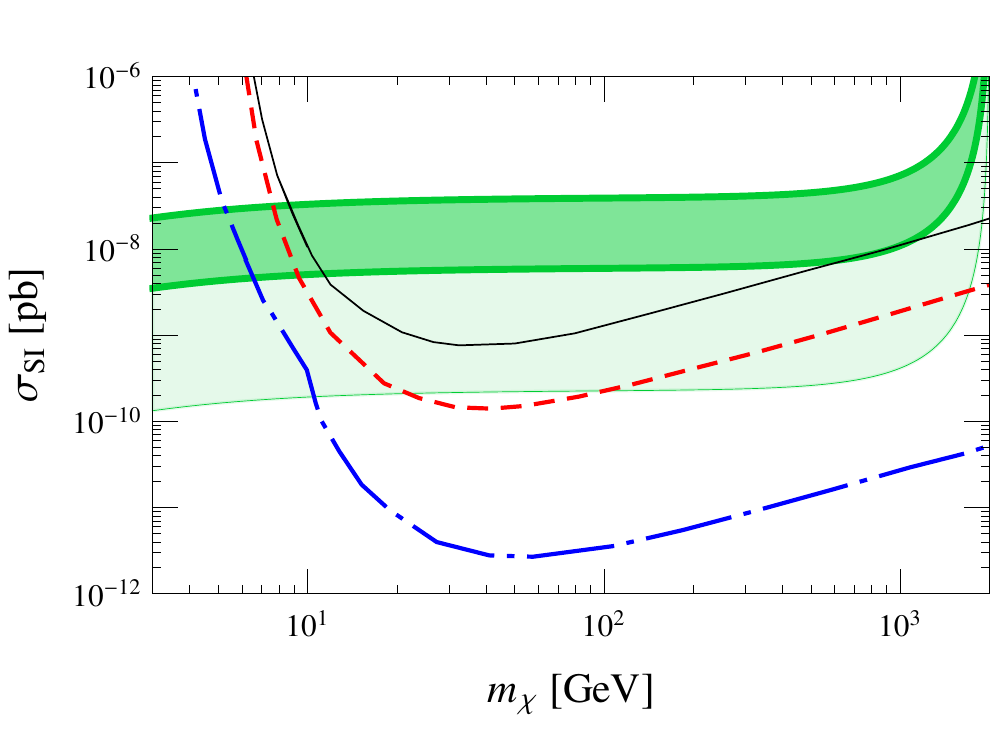}
\end{center}
\caption{{\it Left} Direct detection sensitivity plot in the $(R_s^N, m_{\tilde s_1})$ plane for bino dark matter with $m_{\tilde{\chi}} = 50~\gev$. The grey region (between the plot frame and the solid line) is ruled out by LUX's first data release~\cite{LUX:2013afz}, while the red region (between the solid and dashed lines) was the estimate for 300 days of LUX data, and the blue region (between the dashed and dot-dashed lines) could be probed by LZ~\cite{Cushman:2013zza}. {\it Right} Direct detection sensitivity plots in the $(m_{\tilde{\chi}}, \sigmaSI^N)$ plane for bino dark matter with $m_{\tilde s_1} =2~\tev$ and assuming maximal left-right mixing. The lines have the same meaning as in the left frame. The green bands indicate the uncertainty in the scattering cross section as a function of the strangeness content of the nucleon/$B_q^N$. Appear as Figs.~4 and~5 of Ref.~\cite{Kelso:2014qja}.
}
\label{fig:DDstrange}
\end{figure}

Because  dark matter-nucleon scattering proceeds through the $s$-/$u$-channels, the cross section is enhanced in the limit of small mass splitting between the dark matter and lightest mediator. In the near-degenerate limit, the kinematics of non-relativistic scattering causes the mediator propagator to go nearly on-shell. As a result, next generation direct detection experiments may be sensitive to models well above the mass reach of the LHC in the near-degenerate limit~\cite{Davidson:2017gxx}.

In the scenario where the direct detection cross section is dominated by one light QCD mediator that is significantly heavier than the bino dark matter, the scattering cross section (assuming MSSM-like couplings) will scale with the ratio
\[
R_q^N \equiv Y_{Rq}^2 \sin^2 (2\alpha) (B_q^N)^2\lambda_q^2,
\]
where $Y_{Rq}$ is the hypercharge of the right-handed quark, $B_q^N$ is the integrated nucleon form factor for the quark ($N=p,n$), and $\lambda_q$ accounts for the running of the scattering operator from the weak scale down to the nucleon scale. Assuming maximal mixing,  the left frame of Fig.~\ref{fig:DDstrange} demonstrates that LZ will be sensitive to mediator masses above 5\tev\,for MSSM-like couplings which yield $R_s^N \sim 0.3$.  On the other hand, we see that next generation direct detection experiments can probe models that are consistent with LHC data and that have $\sin (2\alpha) \ll 1$.   

Another interesting result discussed in Ref.~\cite{Kelso:2014qja} and displayed in the right frame of Fig.~\ref{fig:DDstrange} is the very strong dependence of the scattering cross section on the uncertainties in the strangeness content of the nucleon/$B_q^N$ (shown as the green bands) in the circumstance that mediator exchange with strange quarks dominate the scattering.  This uncertainty is reduced to around a factor of 4 if mediator exchange with light quarks dominates the scattering process.  As a note, the most recent results from PandaX-4T \cite{PandaX-4T:2021bab} are about a factor of 2 below the red dashed curve and thus have excluded this particular model at the 90\% confidence level. 

\begin{figure}[t]
\begin{center}
\includegraphics*[width=0.49\textwidth]{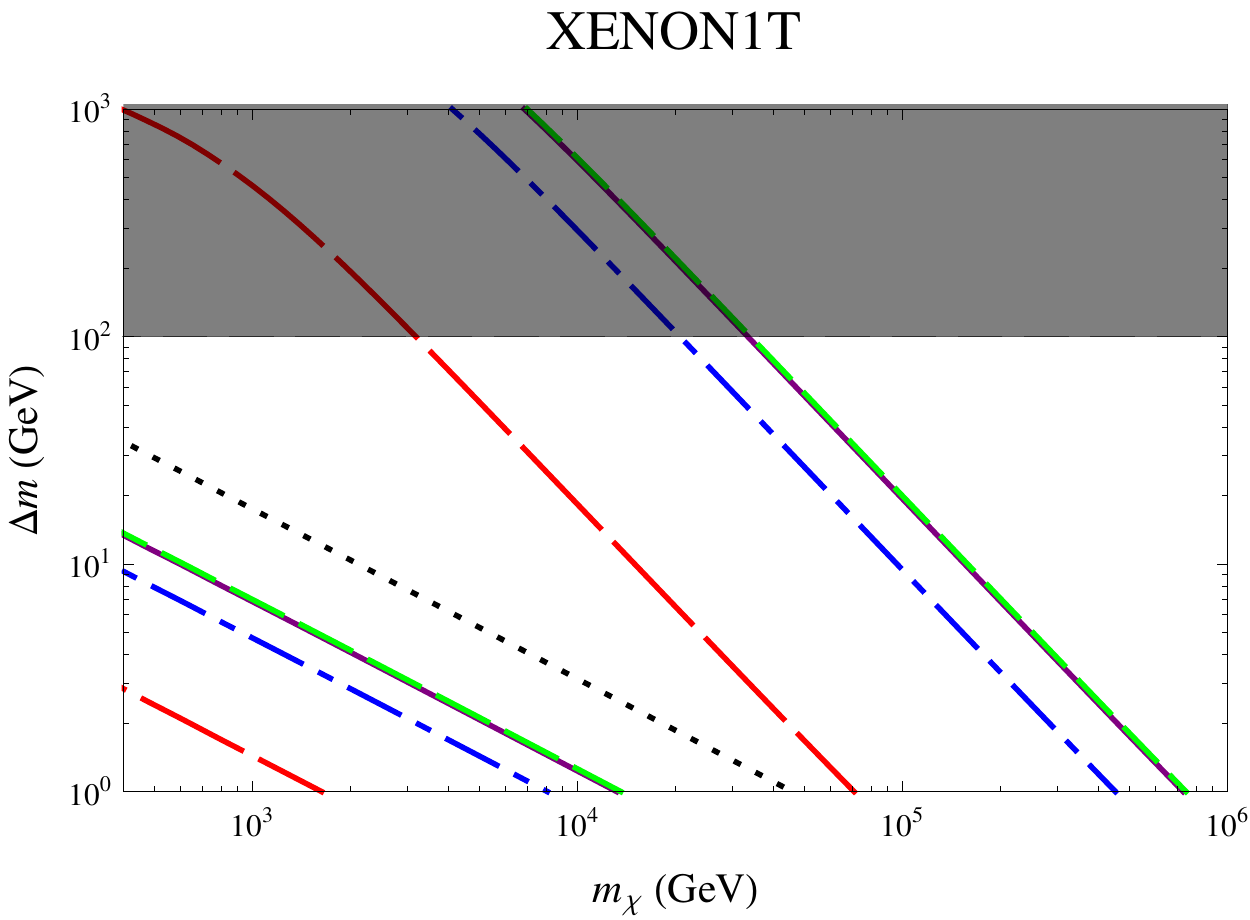}
\includegraphics*[width=0.49\textwidth]{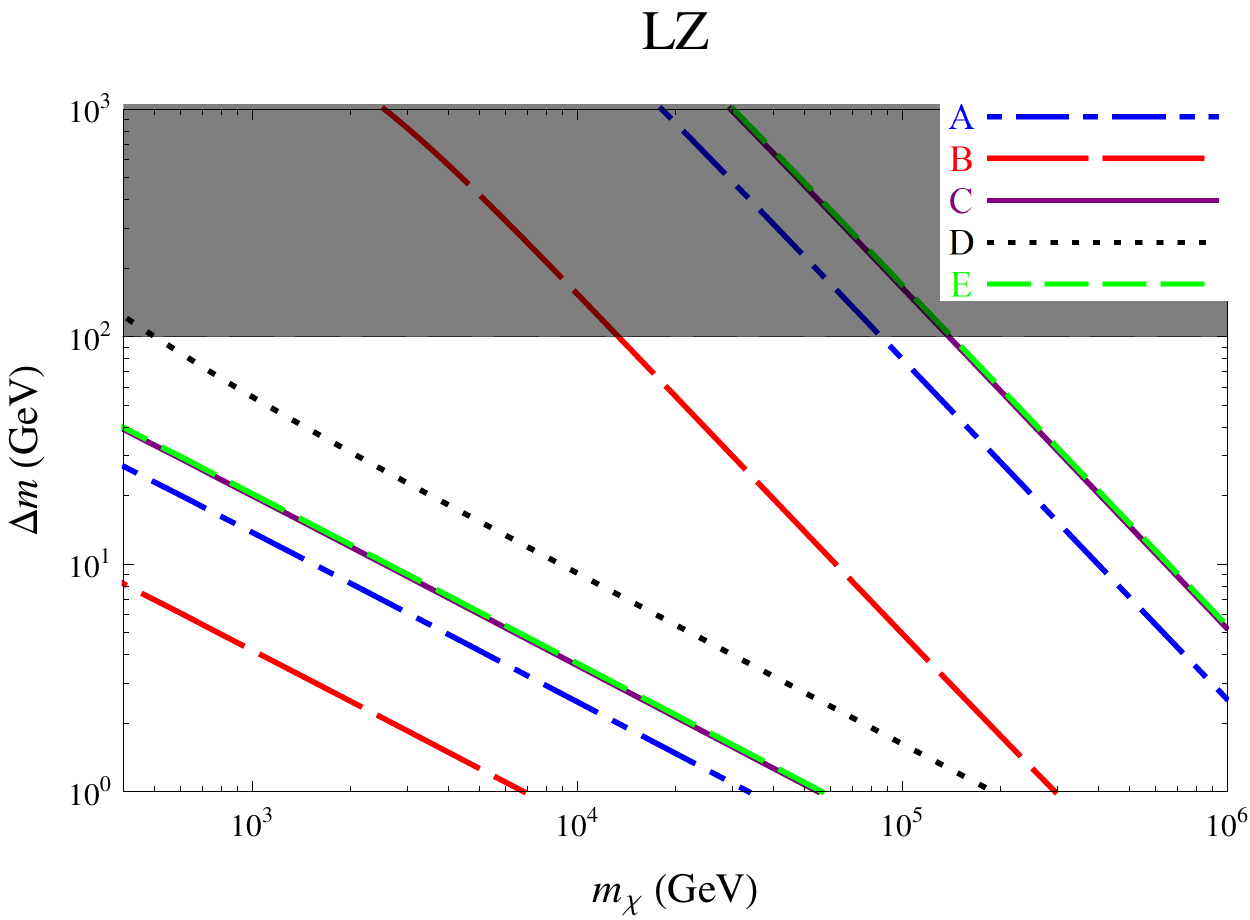}
\end{center}
\caption{Exclusion limit (90\% CL) from XENON1T \cite{XENON:2017vdw} and prospective sensitivity (90\% CL) contours for LZ \cite{Mount:2017qzi} in the $(m_\chi, \Delta m)$-plane for compressed spectra. Note that the most recent PandaX-4T exclusion limits would be about a factor of 2 larger in dark matter mass than the XENON1T results shown.  Note that the parameter space below the contour is excluded. Benchmarks A (dot-dashed blue), B (long-dashed red), C (solid purple), D (dotted black) and E (short-dashed green) are shown. For Benchmarks A, B, C and E, the upper line is the contour if $\alpha = \pi/4$, while the lower line is the contour if $\alpha=0$. For Benchmark D, there is only one contour because the sensitivity is independent of $\alpha$. For the grey shaded region ($\Delta m > 100\gev$), this analysis is not reliable, as the contributions from the heavier squarks cannot be neglected. Appear in Fig.~3 of Ref.~\cite{Davidson:2017gxx}
}
\label{fig:massDiff-LZ}
\end{figure}

As studied in Ref.~\cite{Davidson:2017gxx}, QCD-charged mediators with compressed spectra can lead to significantly enhanced nucleon scattering through a resonance as the propagator of the mediator goes nearly on-shell.  For these compressed spectra, different operators will dominate the nucleon scattering depending on the number of mediators and the chiral mixing angle.  For this reason, several benchmark scenarios are examined:
\begin{itemize}
\item{Benchmark A) a single light squark, $\tilde u_1$;}
\item{Benchmark B) a single light squark, $\tilde s_1$;}
\item{Benchmark C) two light degenerate squarks, $\tilde u_1$ and $\tilde d_1$;}
\item{Benchmark D) two light degenerate squarks, $\tilde u_1$ and $\tilde u_2$;}
\item{Benchmark E) three light degenerate squarks, $\tilde u_1$, $\tilde d_1$ and $\tilde s_1$.}
\end{itemize}
For Benchmarks C, D, and E, the light squarks are assumed to be degenerate, and for all five benchmarks the light squark mass is denoted as $m_{\tilde q}$ and we define $\Delta m \equiv m_{\tilde q} - m_\chi >0$.

Previous exclusion limits \cite{XENON:2017vdw} and future prospects for LZ \cite{Mount:2017qzi} to detect these benchmark scenarios are shown in Fig.~\ref{fig:massDiff-LZ}.  As discussed previously, thermal dark matter can only be consistent with the observed relic abundance if $\Delta m \lsim 25 \gev$ and $m_\chi \lsim 1.5 \tev$. Thus, almost all of the allowed parameter space in Fig.~\ref{fig:massDiff-LZ} requires non-thermal production of the observed relic density.  We observe that even in the limit $\alpha \rightarrow 0$, current direct detection experiments exclude some models for which $m_\chi$ is as large as a few TeV in the degenerate regime. The sensitivity of LZ, in the $\alpha \rightarrow 0$ limit, can extend as far as $\sim 10^5\gev$ (for Benchmark D), and much further for larger mixing angles.  Thus for these models, the sensitivity of future direct detection experiments can far exceed the maximum reach of the LHC, even for small mixing.  In this limit of heavy dark matter whose relic density must be generated non-thermally, direct detection experiments could discover not only dark matter, but also the interactions of QCD-coupled heavy scalars.

The direct detection results presented for the QCD-charged mediators correspond to the choice $\lambda_{Lq,Rq} = \sqrt{2} g' Y_{L,R}$.  One can rescale the sensitivities given above to any other scenario by noting that, at maximal mixing, the DM-nucleus scattering cross section is proportional to $\lambda_L^2 \lambda_R^2$, while for $\alpha =0$ it is proportional to $\lambda_L^4$.

It is also interesting to consider leptonic charged mediators as examined in Refs.~\cite{Sandick:2016zut,Acuna:2021rbg}. In this scenario, scattering with nuclei proceeds only through loop-induced electromagnetic moments. The Majorana nature of our dark matter candidate limits the possibilities to a non-zero anapole moment. The differential scattering cross section for anapole dark matter scattering from the electric field of the nucleus can be written as \cite{Kopp:2014tsa,DelNobile:2014eta,Ho:2012br,Gresham:2013mua}    

\begin{equation}
\label{eq:dsigmadE}
\frac{d\sigma}{dE_{R}}=\alpha_{\text{em}}\mathcal{A}^{2}Z^{2}F^{2}_{E}(\pmb{q}^{2})\left[2m_{T}-\left(1+\frac{m_{T}}{m_{\chi}}\right)^{2}\frac{E_{R}}{v^{2}}\right] \,\,,
\end{equation}
with $\mathcal{A}$ the dark matter anapole moment, $m_{T}$ the mass of the target nucleus, $E_{R}=\pmb{q}^{2}/(2m_T)$ the nuclear recoil energy, $Z$ the nuclear charge, and $F^{2}_{E}$ the nuclear electric form factor. 

\begin{figure}[t]
\begin{center}
\includegraphics*[width=1.0\textwidth]{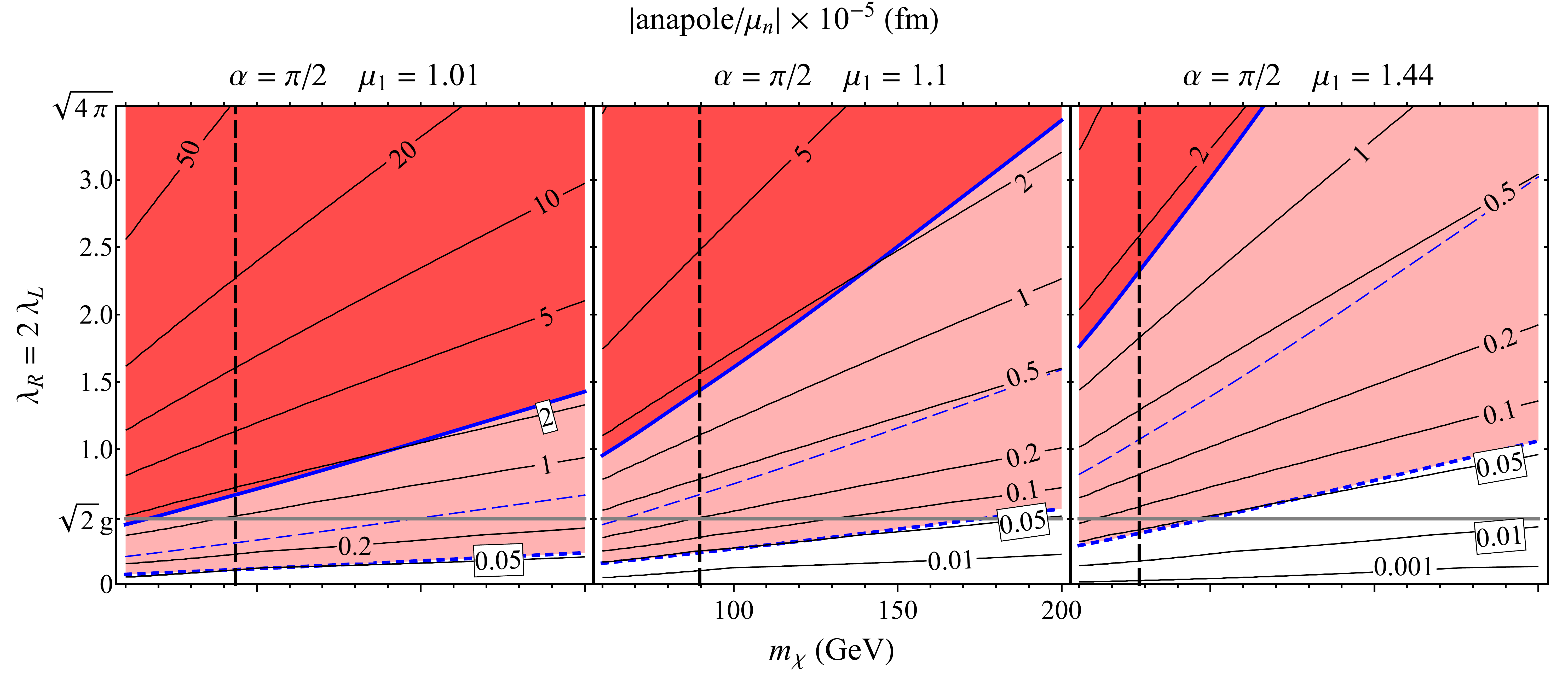}
\end{center}
\caption{We show plots of $\lambda_R = 2 \lambda_L$ versus $m_\chi$ for $\mu_1 = 1.01$ (left), $\mu_1 = 1.10$ (center), and $\mu_1 = 1.44$ (right) for maximal mixing $(\alpha = \pi/2)$. The contours correspond to values of $|\mathcal{A}/\mu_N| \times 10^{-5}$ fm. The vertical black dashed line is the LEP limit on the mass of the charged scalar for the $\mu$ channel.  The solid (dashed) blue lines correspond to LUX 2014 (future LZ) limits on the DM SI scattering cross section.  In addition, the most recent LUX 2016 constraint is estimated as a thin dashed blue contour.  The gray horizontal line corresponds to the SUSY value of couplings. Appear in Fig.~6 of Ref.~\cite{Sandick:2016zut}
}
\label{fig:anapoleDM}
\end{figure}

We present the most optimistic limits and projections for leptonic charged mediators for the $\mu$ channel in Fig.~\ref{fig:anapoleDM}.  The three different panels correspond to different levels of degeneracy of the mass of the mediator as characterized by $\mu_{1}=m^{2}_{\widetilde{f}_{1}}/m_{\chi}^{2}$. The solid (dashed) blue lines correspond to LUX 2014 \cite{LUX:2015abn} (future LZ \cite{LZ:2015kxe}) limits on the DM SI scattering cross section.  In addition, the LUX 2016 constraint \cite{LUX:2016ggv} is estimated as a thin dashed blue contour. We observe that for $\alpha=\pi/2$, LZ will probe a significant portion of the parameter space of these models for $m_\chi \sim \mathcal{O}(100-200)$ GeV and lightest mediator mass within $\mathcal{O}(5\%)$ of the DM mass. SUSY-level couplings can be tested out to DM mass of 100 GeV with the lightest mediator mass within $\mathcal{O}(20\%)$ of the DM mass. Dialing the mixing angle down to zero significantly weakens direct detection prospects.   Although not shown here, Ref.~\cite{Sandick:2016zut} demonstrates that SUSY-level couplings can be tested by LZ only out to DM masses of 90~GeV with the lightest mediator mass within $\mathcal{O}(5\%)$ of the DM mass.

\subsection{Indirect detection}

In the charged mediator models discussed in this white paper, the most straight-forward avenue for dark matter indirect detection is through the annihilation process $\s{B}\s{B} \to \bar f f$.  If the thermal relic density is depleted in the early Universe by $s$-wave annihilation, then this process will be dominant in the current epoch. In particular, dark matter annihilation to hadronic final states or taus (which can subsequently decay hadronically) can yield detectable signals at gamma-ray telescopes. Fermi-LAT searches for dark matter annihilation in dwarf spheroidals (dSphs)~\cite{Fermi-LAT:2016uux} constrain this scenario, but for $m_{\s{B}} \gtrsim {\cal O}(100~\gev)$, there is allowed parameter space in which the $s$-wave annihilation cross section is consistent with the relic density. But future data, as well as improvements in $J$-factor determinations, may more tightly probe this regime.  

\begin{figure}
\begin{center}
\includegraphics[width=0.49\hsize]{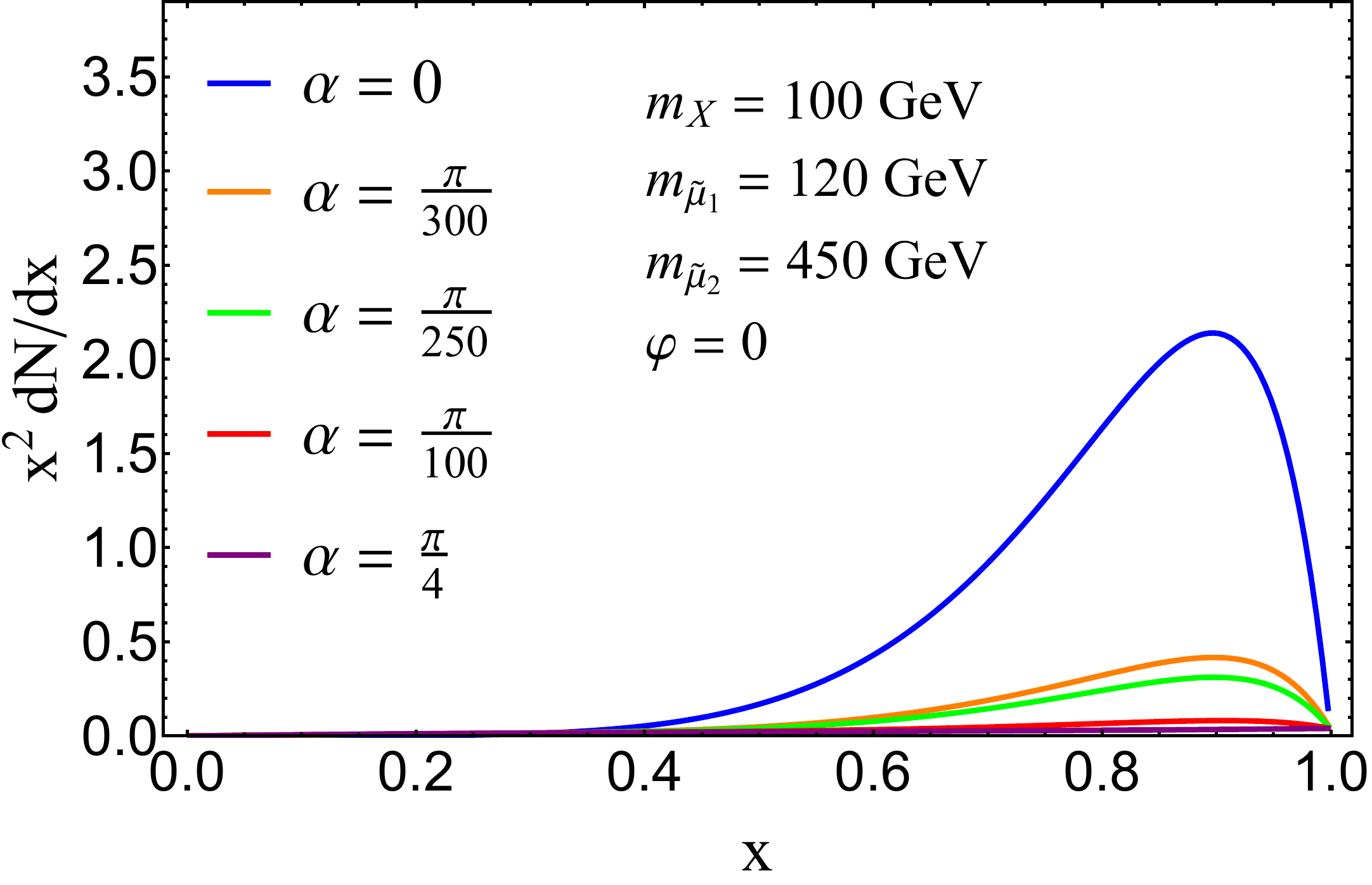}
\includegraphics[width=0.49\hsize]{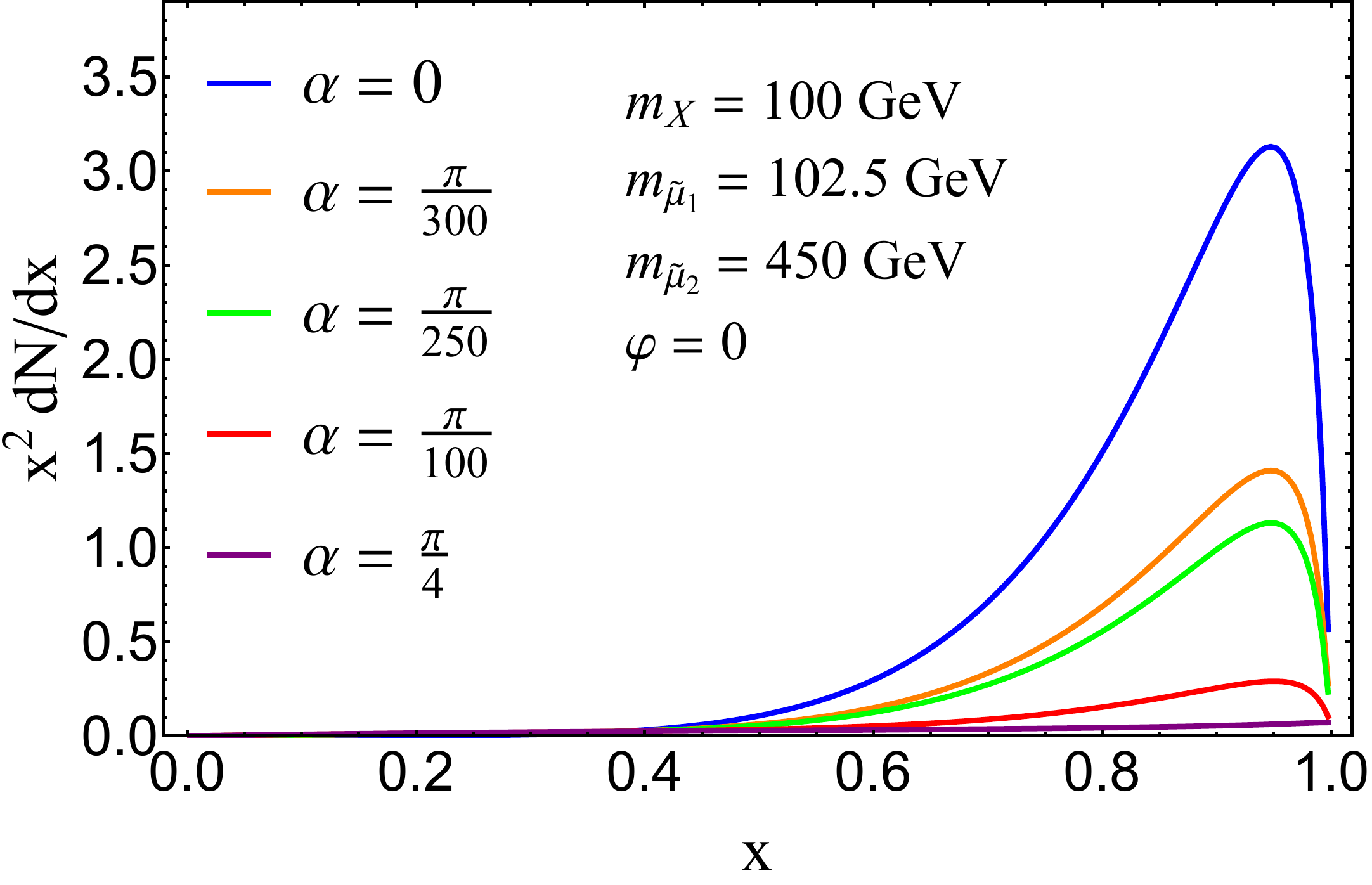}
\end{center}
\caption{Photon spectra produced by $X X\to
\bar f f \gamma$ for two benchmark mass splittings between the bino and lightest smuon, assuming a range of left-right mixing angles, $\alpha$, and MSSM-like couplings, $\lambda_L = (\sqrt{2}/2) g$ and $\lambda_R = \sqrt{2} g$. Appears in Fig.~8 of Ref.~\cite{Kumar:2016cum}.
}
\label{fig:IBspec}
\end{figure}

If dark matter annihilates to muons, this scenario is more difficult to probe with gamma-ray telescopes, since muon decay produces few photons. However, as discussed in Ref.~\cite{Kumar:2016cum}, there are alternative gamma-ray signals which can be important in this case. For example, the process $\s{B}\s{B} \to \gamma \gamma$ can proceed at one-loop, resulting in a striking line signal. The photon spectrum from bino annihilation can be written as a sum of contributions from line signals and continuum emission,
\begin{equation}
    \frac{dN}{dx} = \left( \frac{dN}{dx} \right)_{\rm line} + \left( \frac{dN}{dx} \right)_{\rm cont.} \, ,
\end{equation}
with $x = E_\gamma/ m_{\s{B}}$. We will focus on the constraints from continuum emission, 
\begin{equation}
    \left( \frac{dN}{dx} \right)_{\rm cont.} = \frac{1}{\VEV{\sigma_{\s{B}\s{B}} v}} \left[ \frac{d \VEV{\sigma_{\rm IB} v}}{dx} + \sum_i N_i \frac{d \VEV{\sigma_i v}}{dx} \right] \, ,
\end{equation}
where the leading contribution is from internal bremsstrahlung (IB) $\s{B}\s{B} \to \bar f f \gamma$ and the second term is a sum over higher order processes which each yield $N_i$ photons in a single annihilation. 

In the limit of small final state fermion mass $m_f / m_{\s{B}} \to 0$ and negligible left-right sfermion mixing angle, the dominant contribution to the IB cross section is given by  Eq.~\leqn{eq:IB} of Appendix~\ref{app:IB}. Although the cross section is suppressed relative to that for $\s{B}\s{B}\to \bar f f$ by an additional factor of $\alpha_{em}$, it is neither $p$-wave nor chirality-suppressed. Moreover, the associated contribution to the cross section in Eq.~\leqn{eq:IB} from virtual internal bremsstrahlung (VIB) often results in the production of a hard photon, which can look almost line-like, yielding a much more easily detectable signal. 

In Fig.~\ref{fig:IBspec}, we plot the total IB photon spectra for different bino-smuon mass splittings and left-right mixing angles. We see the line-like feature at $E_\gamma \simeq m_X$ is more prominent for smaller mass splittings. As the left-right mixing angle increases the VIB contribution to the IB amplitude eventually becomes subdominant to the contribution associated with final state radiation (FSR). Since the FSR contribution is enhanced for soft and collinear photons, the associated IB photon spectra become relatively flat in energy.

\begin{figure}
\begin{center}
\includegraphics[width=0.49\hsize]{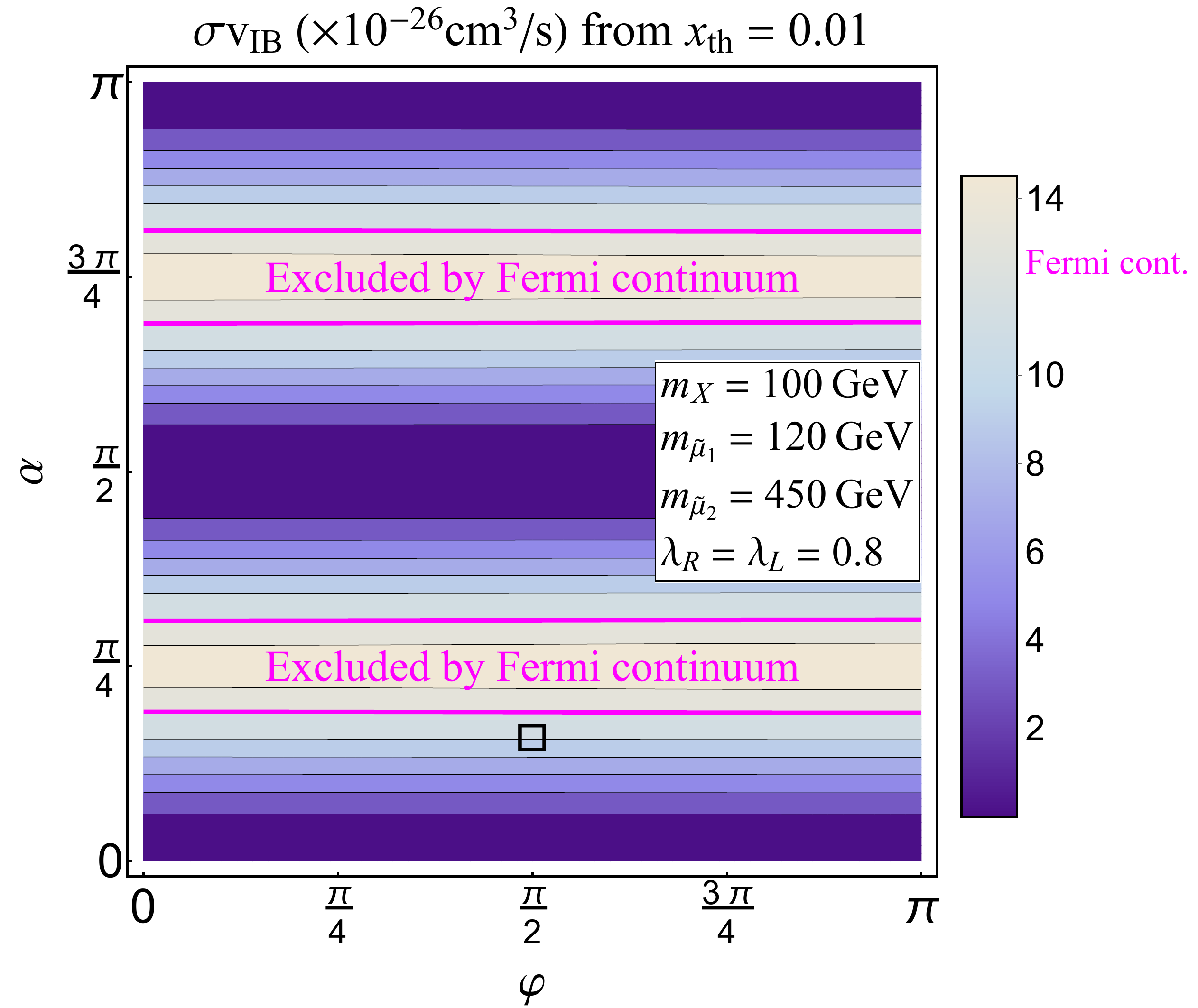}
\includegraphics[width=0.49\hsize]{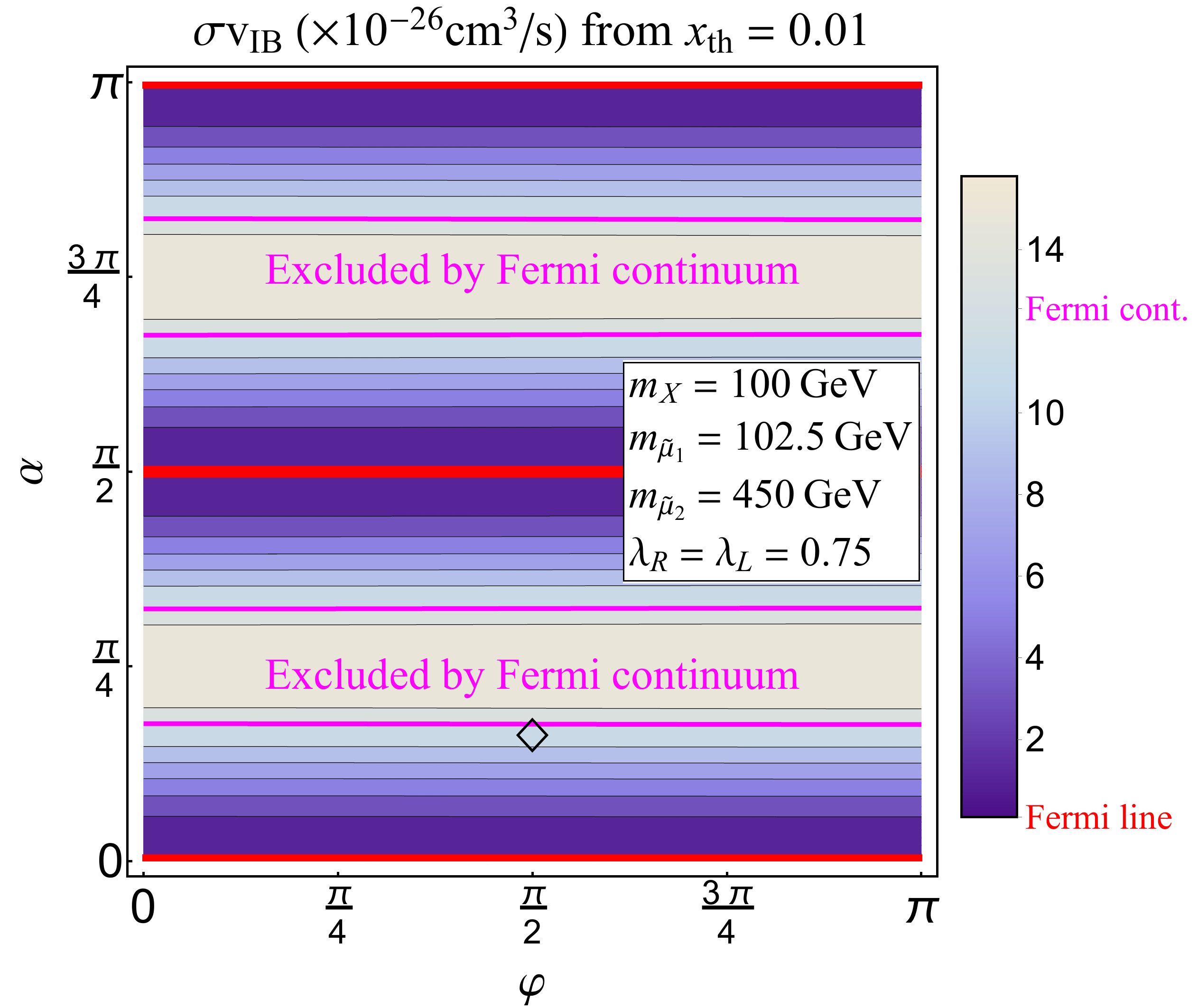}
\end{center}
\caption{Contours showing the internal bremsstrahlung (IB), $X X \to \bar f f \gamma$, cross section for models with larger (smaller) bino-smuon mass splitting in the left (right) panel. Given the corresponding gamma ray spectra shown for similar benchmark models in Fig.~\ref{fig:IBspec}, we also indicate the parameter space of these models constrained by Fermi-LAT limits on continuum photon emission~\cite{Geringer-Sameth:2011wse} and gamma-ray line searches~\cite{Fermi-LAT:2015kyq}. Appears as Fig.~10 of Ref.~\cite{Kumar:2016cum}.
}
\label{fig:IBconst}
\end{figure}

To estimate the sensitivity of Fermi-LAT searches for dark matter annihilation in dSphs to continuum signals from IB in charged mediator models, we can integrate the photon spectra to calculate the particle physics factor~\cite{Geringer-Sameth:2011wse},
\begin{equation}
    \Phi_{\rm PP} = \frac{\VEV{\sigma_{\s{B}\s{B}} v}}{8 \pi m_{\s{B}}^2} \int_{x_{\rm th}}^1 dx \, \left( \frac{dN}{dx} \right)_{\rm cont.} .
\end{equation}
Although the analysis in Ref.~\cite{Geringer-Sameth:2011wse} does not incorporate the most recent Fermi-LAT data, the limit on $\Phi_{\rm PP} < 5.0_{-4.5}^{+4.3} \times 10^{-30} \, {\rm cm^3 s^{-1}} \GeV^{-2}$ can be used to constrain the continuum photon signal from IB for the variety of spectral shapes shown in Fig.~\ref{fig:IBspec}.  Indeed, more current limits from Fermi-LAT dSph searches on models with charged mediators that couple to leptons have been presented, {\it e.g.}, in Ref.~\cite{Boddy:2018qur}.

For models described in Sec.~\ref{sec:bulk} which can satisfy the dark matter relic density through bino annihilation, the IB signal is typically too small assuming the MSSM benchmark couplings defined after Eq.~\ref{eq:couplings}. However, if we consider the parameter space of a more general simplified model, then Fermi-LAT could be sensitive to the IB signal from dSphs. In Fig.~\ref{fig:IBconst}, we show contours of the IB cross section for models with the same mass spectra as in Fig.~\ref{fig:IBspec}, but with $|\lambda_{L,R}| \sim 1$. We see that the Fermi limit on the continuum emission is strongest towards the limit of maximal left-right mixing, where the FSR contribution dominates the IB signal. In the opposite limit with small left-right mixing, the line-like feature becomes more prominent as the VIB contribution becomes dominant and, as can be seen in the right panel for smaller bino-smuon mass splitting, can be constrained by Fermi-LAT gamma-ray line searches~\cite{Fermi-LAT:2015kyq}.  

While we have focused on gamma-ray signals from dSphs, there are other potentially interesting indirect detection signatures associated with charged mediator models. For example, AMS-02 searches for positrons produced in DM annihilation can yield stringent limits ($\lsim \pb$) on the cross section for the process $\s{B} \s{B} \to \mu^+ \mu^-$~\cite{Bergstrom:2013jra}. In principal, these constraints from cosmic ray experiments are in tension with several of the benchmark models discussed above, particularly those with couplings $|\lambda_{L,R}| \sim 1$ shown in Fig.~\ref{fig:IBconst}. However, positron searches are subject to relatively large systematic uncertainties associated with cosmic ray propagation and astrophysical background modeling (for example, see~\cite{DiMauro:2015jxa}). Alternatively, even for models with velocity-suppressed $\s{B} \s{B} \to \mu^+ \mu^-$ cross sections, gamma-ray signals could be significant for bino annihilation in dark matter spikes formed around supermassive black holes, where the dark matter velocity dispersion can be enhanced~\cite{Sandick:2016zeg}.

\section{Conclusion} \label{sec:con}

In this white paper, we describe the basic aspects of dark matter models with charged mediators. For a Majorana singlet dark matter candidate with a mass between $100 \, \GeV $ and $1 \, \TeV$, interactions with the Standard Model are mediated by scalar partners of leptons and quarks. Although the MSSM is a well-studied extension to the Standard Model with charged mediators, we stress that the simplified models considered in the work summarized by this white paper constitute a more general framework with cosmologically viable dark matter production mechanisms and novel phenomenological signatures.

For models with $\O(100) \, \GeV$ scalar fermion partners, the observed relic density can be satisfied through $s$- or $p$-wave dark matter annihilation. The significant left-right mixing necessary for the scalar mediators is usually not considered in typical realizations of the MSSM which incorporate minimal flavor violation. Significant intragenerational mixing can, in turn, yield a variety of potentially interesting experimental signatures. For instance, while stringent limits on the dipole moments of the electron severely constrain scenarios with first generation scalar leptons, models with scalar muon partners in this ``Incredible Bulk" scenario can account for the $g_\mu -2$ anomaly in addition to satisfying the relic density.

Further, while current LHC constraints on the masses of first and second generation scalar lepton partners have only made marginal improvements compared to LEP for compressed spectra, we discuss new search strategies which can potentially improve the sensitivity of LHC to such models. We also summarize the sensitivity of indirect detection searches, focusing on the gamma-ray signal from internal bremsstrahlung in models with scalar muons mediating dark matter annihilation. For models with either scalar lepton partners or QCD-charged scalar mediators, we both describe the current constraints from and project the future sensitivity of direct detection experiments.

In models with scalar fermion partners nearly degenerate in mass with the dark matter candidate can the relic density can, alternatively, be depleted through co-annihilation processes involving scalar mediators with masses up to $\sim 1 \, \TeV$. Both direct and indirect detection signatures can be enhanced in models with compressed spectra. Particularly relevant for future dark matter searches, models with QCD-charged scalars mediating interactions between dark matter and nuclei could be probed by direct detection experiments for scalar masses well beyond the reach of the LHC. For models with scalar lepton partners at the $\TeV$ scale, we demonstrate how a future lepton collider could probe the parameter space relevant for dark matter co-annihilation and $g_\mu -2$.

There are several potentially interesting avenues for future work on the phenomenology of charged mediator models. Although many of the models we have discussed can be difficult to detect at direct detection experiments, dark matter capture in compact objects could probe models with velocity- or momentum-suppressed dark matter-nucleon interactions. Also, the new search strategies we propose for scalar lepton partner searches at LHC could be further optimized in analyses which utilize machine learning techniques. Regarding models with scalar lepton partners which satisfy the relic density through co-annihilation, stringent constraints from perturbative unitarity and vacuum stability warrant further study of how these simplified models might be embedded in a more complete extension of the Standard Model.

\Acknowledgements

The authors are grateful for the collaboration of Jan Tristram Acu\~{n}a, Sebastian Baum, Andrew Davidson, Bhaskar Dutta, Kebur Fantahun, Ashen Fernando, Keita Fukushima, Kuver Sinha, Fei Teng, Piero Ullio, Joel W. Walker and Takahiro Yamamoto on the works summarized in this white paper. The work of T.~Ghosh is supported by DAE, India for the Regional Centre for Accelerator based Particle Physics (RECAPP), Harish Chandra Research Institute. The work of J.~Kumar is supported in part by DOE grant DE-SC0010504. The work of P.~Sandick is supported by NSF grant PHY-2014075. The work of P.~Stengel is partially supported by the research grant ``The Dark Universe: A Synergic Multi-messenger Approach" number 2017X7X85K under the program PRIN 2017 funded by the Ministero dell'Istruzione, Universit\`{a} e della Ricerca (MIUR), and by the ``Hidden" European ITN project (H2020-MSCA-ITN-2019//860881-HIDDeN). 

\appendix
\section{Relic density} \label{app:relic}
\subsection{Bino annihilation cross section} \label{app:bulk}
Assuming the MSSM benchmark for the Yukawa couplings $\lambda_{L,R}$, the velocity-independent $s$-wave coefficient in Eq.~\leqn{eqn:annx} is given by
\begin{equation} \label{eq:c0}
    c_0 = \frac{m_{\s{B}}^2}{2 \pi} g^4 Y_L^2 Y_R^2 \cos^2 \alpha \sin^2 \alpha \left(\frac{1}{m_{\s{f}_1}^2 + m_{\s{B}}^2} - \frac{1}{m_{\s{f}_2}^2 + m_{\s{B}}^2} \right)^2 \, ,
\end{equation}
where the masses of the charged mediator mass eignestates are $m_{\s{f}_{1,2}}$, and the coefficient for the first velocity suppressed term is given by
\begin{align} \label{eq:c1}
    c_1 = \frac{m_{\s{B}}^2}{2 \pi} g^4  
    &\Biggl(
    \frac{( Y_L^4 \cos^4 \alpha + Y_R^4 \sin^4 \alpha )( m_{\s{f}_1}^4 + m_{\s{B}}^4 )}{(m_{\s{f}_1}^2 + m_{\s{B}}^2 )^4} +
    \frac{( Y_L^4 \sin^4 \alpha + Y_R^4 \cos^4 \alpha )( m_{\s{f}_2}^4 + m_{\s{B}}^4 )}{(m_{\s{f}_2}^2 + m_{\s{B}}^2 )^4} 
    \nonumber \\ 
    & + \frac{2 (Y_L^4 + Y_R^4) \sin^2 \alpha \cos^2 \alpha (m_{\s{f}_1}^2 m_{\s{f}_2}^2 + m_{\s{B}}^4)}{(m_{\s{f}_1}^2 + m_{\s{B}}^2 )^2 (m_{\s{f}_2}^2 + m_{\s{B}}^2 )^2} 
    \nonumber \\ 
    & + \frac{Y_L^2 Y_L^2 \sin^2 \alpha \cos^2 \alpha (m_{\s{f}_1}^2 - m_{\s{f}_2}^2)^2}{2 (m_{\s{f}_1}^2 + m_{\s{B}}^2 )^4 (m_{\s{f}_2}^2 + m_{\s{B}}^2 )^4} \left[ 3 m_{\s{f}_1}^4 m_{\s{f}_2}^4 - 52 m_{\s{B}}^4 m_{\s{f}_1}^2 m_{\s{f}_2}^2 + 3 m_{\s{B}}^8
    \right. \nonumber \\
    & \left. 
    -14 m_{\s{B}}^2 (m_{\s{f}_1}^2 + m_{\s{f}_2}^2 )(m_{\s{B}}^4 + m_{\s{f}_1}^2 m_{\s{f}_2}^2) - 5 m_{\s{B}}^4 (m_{\s{f}_1}^4 + m_{\s{f}_2}^4 )
    \right] \Biggr) \, .
\end{align}
In the above coefficients, we have assumed all terms proportional to the SM fermion mass $m_f$ are small. While this approximation holds reasonably well for models with first and second generation scalar mediators, contributions from terms $\propto m_f$ can be significant for models with the bino coupled to scalar partners of third generation SM fermions. 

\subsection{Squark annihilation cross section} \label{app:coann}
The contributions to the effective annihilation cross section from squark annihilation to gluons in Eq.~\leqn{eqn:coannx} can be summed over light quark flavors,
\begin{equation} \label{eqn:sqsqx}
    \sum_{q = u,d,s,c} \frac{n^{\rm eq}_{\s{q}} n^{\rm eq}_{\s{q}}}{\left[n^{\rm eq} \right]^2} \VEV{\sigma_{\s{q} \s{q}} v} = \frac{7 g_s^4 N_{\s{q}}}{432 \pi m_{\s{q}}^2} \left[ 
    N_{\s{q}} + \frac{\exp(\Delta m / T)}{3(1+\Delta m / m_{\s{B}})^{3/2}}
    \right]^{-2} \, ,
\end{equation}
for the QCD gauge coupling $g_s$ and assuming $N_{\s{q}}$ mass degenerate squarks with a mass splitting relative to the bino $\Delta m = m_{\s{q}} - m_{\s{B}}$.
  
\section{Phenomenology}  \label{app:pheno}

\subsection{Fermion dipole moment corrections} \label{app:dip}

If $m_f \ll m_\chi$, then one-loop diagrams with dark matter and the charged mediators in the loop yield contributions to the anomalous magnetic moment ($a = (g-2) /2 $) and to the electric dipole moment ($d/|e|$) of the SM fermion, given by~\cite{Cheung:2009fc}
\bea
\Delta a &=&
 {m_{\ell} m_{\tilde \chi} \over 4 \pi^2 m_{\tilde f_1}^2} g^2 Y_L Y_R \cos \varphi
 \cos \alpha \sin \alpha   \left[ {1 \over 2(1- r_{1})^2} \left( 1+  r_{1} + {2 r_{1} \ln  r_{1} \over
 1- r_{1}}\right)   \right]
 -(\tilde f_1 \to \tilde f_2)
\nonumber \\
 {d \over |e|} &=& {m_{\tilde \chi} \over 8 \pi^2 m_{\tilde f_1}^2 } g^2 Y_L Y_R \sin \varphi
 \cos \alpha \sin \alpha  \left[ {1 \over 2(1- r_{1})^2} \left( 1+  r_{1} + {2 r_{1} \ln  r_{1} \over
 1- r_{1}}\right)   \right]
\nonumber \\  &&
-(\tilde f_1 \to \tilde f_2)
\label{eq:DMs}
\eea
where $ r_{i} \equiv m_{\s{B}}^2/m_{\s{f}_i}^2$. While the contributions to both the magnetic and electric dipole moments require non-vanishing left-right mixing angle $\alpha$, increasing the $CP$-violating phase $\phi$ from $0$ to $\pi/2$ suppresses $\Delta a$ and enhances $d/|e|$.

\subsection{Effective interactions for direct detection}
\label{app:DD}

Dark matter scattering with quarks via exchange of the charged mediators in the $s$-/$u$-channel can be expressed in terms of effective contact operators
\bea
{\cal O}_q &=& \sum_{i=1}^7 {\cal O}_{qi},
\eea
where the dimension-6 operators are (using the notation of ~\cite{Fukushima:2011df})
\bea
{\cal O}_{q1} &=& \alpha_{q1} (\bar \chi \gamma^\mu \gamma^5 \chi) (\bar q \gamma_\mu q) ,
\nonumber\\
{\cal O}_{q2} &=& \alpha_{q2} (\bar \chi \gamma^\mu \gamma^5 \chi) (\bar q \gamma_\mu \gamma^5 q) ,
\nonumber\\
{\cal O}_{q3} &=& \alpha_{q3} (\bar \chi  \chi) (\bar q  q) ,
\nonumber\\
{\cal O}_{q4} &=& \alpha_{q4} (\bar \chi \gamma^5 \chi) (\bar q  \gamma^5 q) ,
\nonumber\\
{\cal O}_{q5} &=& \alpha_{q5} (\bar \chi \chi) (\bar q  \gamma^5 q) ,
\nonumber\\
{\cal O}_{q6} &=& \alpha_{q6} (\bar \chi \gamma^5 \chi) (\bar q   q) ,
\eea
with
\bea
\alpha_{q1} &= &
-\left[{|\lambda_L^2| \over 8}
\left( {\cos^2 \alpha \over m_{\tilde q_1}^2 -m_\chi^2}
+ {\sin^2 \alpha \over m_{\tilde q_2}^2 -m_\chi^2} \right)
-{|\lambda_R^2| \over 8}
\left({\cos^2 \alpha \over m_{\tilde q_2}^2 -m_\chi^2}
+ {\sin^2 \alpha \over m_{\tilde q_1}^2 -m_\chi^2} \right) \right] ,
\nonumber\\
\alpha_{q2} &= &
\left[{|\lambda_L^2| \over 8}
\left( {\cos^2 \alpha \over m_{\tilde q_1}^2 -m_\chi^2}
+ {\sin^2 \alpha \over m_{\tilde q_2}^2 -m_\chi^2} \right)
+{|\lambda_R^2| \over 8}
\left({\cos^2 \alpha \over m_{\tilde q_2}^2 -m_\chi^2}
+ {\sin^2 \alpha \over m_{\tilde q_1}^2 -m_\chi^2} \right) \right] ,
\nonumber\\
\alpha_{q3,4} &=& {Re(\lambda_L \lambda_R^*) \over 4}(\cos \alpha \sin \alpha)
\left[{1\over m_{\tilde q_1}^2 -m_\chi^2} -{1\over m_{\tilde q_2}^2 -m_\chi^2} \right] ,
\nonumber\\
\alpha_{q5,6} &=& {\imath Im(\lambda_L \lambda_R^*) \over 4}(\cos \alpha \sin \alpha)
\left[{1\over m_{\tilde q_1}^2 -m_\chi^2} -{1\over m_{\tilde q_2}^2 -m_\chi^2} \right] .
\label{eq:Coefficients}
\eea

One obtains higher dimension contact operators by  expanding the scalar propagators in powers momenta.  These operators are generally subdominant because their contributions to the dark matter-nucleon scattering matrix element are suppressed by additional factors of $ \sim (m_N/ 2 (m_{\tilde q_1} - m_\chi))$. However, a dimension-8 twist-2 operator can be significant~\cite{Drees:1993bu}:
\bea
{\cal O}_{q7} &=& \alpha_{q7 } (\imath \bar \chi \gamma^\mu \partial^\nu \chi )
\left[ \left(\frac{\imath}{2} \right)
\left(\bar q \gamma_\mu \partial_\nu q + \bar q \gamma_\nu \partial_\mu q - \frac{1}{2} g_{\mu \nu} \bar q \gamma_\alpha \gamma^\alpha q  \right) \right] ,
\eea
where
\bea
\alpha_{q7} &=&
\frac{| \lambda_L^2 |}{4} \left[ { \cos^2 \alpha \over (m_{\tilde{q}_1}^2 - m_\chi^2)^2 }
+ { \sin^2 \alpha \over (m_{\tilde{q}_2}^2 - m_\chi^2)^2 } \right]
+ \frac{| \lambda_R^2 |}{4}
\left[ { \cos^2 \alpha \over (m_{\tilde{q}_2}^2 - m_\chi^2)^2 }
+ { \sin^2 \alpha \over (m_{\tilde{q}_1}^2 - m_\chi^2)^2 } \right] .
\eea
This operator can generate an important contribution to the SI velocity-independent scattering matrix element in the limit of small mixing ($\alpha \rightarrow 0$), especially if $(m_{\tilde q_1} - m_\chi) / m_\chi \ll 1$.

\subsection{Internal bremsstrahlung cross section} \label{app:IB}
  
The contribution to the differential cross section for $\s{B} \s{B} \to \bar f f \gamma$ associated with the production of fermion-antifermion pairs with opposite helicity in the $m_f \to 0$ limit is~\cite{Kumar:2016cum}
\begin{align} \label{eq:IB}
  \frac{d \VEV{\sigma_{\rm vb} v}}{dx} = \sum_{i=1,2} \frac{\alpha_{\rm em} \lambda_i^4 (1-x)}{64 \pi^2 m_{\s{B}}^2}
  & \left[   
  \frac{4 x}{(1+\mu_i)(1+\mu_i - 2 x)} - \frac{2x}{(1+\mu_i - x)^2}  
  \right. \nonumber \\
  & \left. - \frac{(1+\mu_i)(1+\mu_i - 2 x)}{(1+\mu_i - x)^3} 
  \log \frac{1+\mu_i}{1+\mu_i - 2 x} 
  \right] \, , 
\end{align}
where
\begin{align}
 \lambda_1^2 = | \lambda_L |^2 \cos^2 \alpha - | \lambda_R|^2 \sin^2 \alpha \, , 
 \nonumber \\
 \lambda_2^2 = | \lambda_L |^2 \sin^2 \alpha - | \lambda_R|^2 \cos^2 \alpha \, , \nonumber
\end{align}
$\mu_i = m_{\s{f}_i}^2 / m_{\s{B}}^2$ and $\alpha_{\rm em}$ is the electromagnetic coupling constant. For sfermions nearly degenerate with the bino, $\mu_i \sim 1$ and the differential cross section becomes strongly peaked near $x \sim 1$. This corresponds to the limit where one of the final state fermions becomes soft and a sfermion propagator goes on-shell.





\bibliographystyle{JHEP}
\bibliography{charged_mediators}  


\end{document}

%% file: workshopsymbols.tex
\def\ie{{\it i.e.}}
\def\eg{{\it e.g.}}

\def\Acknowledgements{\bigskip  \bigskip \begin{center} \begin{large}
             \bf ACKNOWLEDGEMENTS \end{large}\end{center}}

\def\beq{\begin{equation}}
\def\eeq#1{\label{#1}\end{equation}}
\def\eeqn{\end{equation}}

\newenvironment{Eqnarray}%
   {\arraycolsep 0.14em\begin{eqnarray}}{\end{eqnarray}}
\def\beqa{\begin{Eqnarray}}
\def\eeqa#1{\label{#1}\end{Eqnarray}}
\def\eeqan{\end{Eqnarray}}

\def\leqn#1{(\ref{#1})}

\let\bar=\overbar

\def\VEV#1{\left\langle{ #1} \right\rangle}

\def\lsim{\mathrel{\raise.3ex\hbox{$<$\kern-.75em\lower1ex\hbox{$\sim$}}}}
\def\gsim{\mathrel{\raise.3ex\hbox{$>$\kern-.75em\lower1ex\hbox{$\sim$}}}}

\def\L{{\cal L}}

\def\O{{\cal O}}

\def\del{\partial}
\def\Dslash{\not{\hbox{\kern-4pt $D$}}}
\def\dslash{\not{\hbox{\kern-2pt $\del$}}}
\def\pslash{\not{\hbox{\kern-2pt $p$}}}
\def\ETmiss{\not{\hbox{\kern-4pt $E$}}_T}

\def\Dlr{\mathrel{\raise1.5ex\hbox{$\leftrightarrow$\kern-1em\lower1.5ex\hbox{$D$}}}}

\def\ee{e^+e^-}

\def\MSB{{\bar{M \kern -2pt S}}}
\def\msb{{\bar{\scriptsize M \kern -1pt S}}}

\def\drb{{\bar{\scriptsize D \kern -1pt R}}}

\def\GeV{{\rm GeV}}
\def\TeV{{\rm TeV}}

\def\s#1{\widetilde{#1}}


%% file: charged_mediators.bbl
\providecommand{\href}[2]{#2}\begingroup\raggedright\begin{thebibliography}{10}

\bibitem{Fukushima:2014yia}
K.~Fukushima, C.~Kelso, J.~Kumar, P.~Sandick and T.~Yamamoto, \emph{{MSSM dark
  matter and a light slepton sector: The incredible bulk}},
  \href{https://doi.org/10.1103/PhysRevD.90.095007}{\emph{Phys. Rev. D}
  {\bfseries 90} (2014) 095007}
  [\href{https://arxiv.org/abs/1406.4903}{{\ttfamily 1406.4903}}].

\bibitem{Davidson:2017gxx}
A.~Davidson, C.~Kelso, J.~Kumar, P.~Sandick and P.~Stengel, \emph{{Study of
  dark matter and QCD-charged mediators in the quasidegenerate regime}},
  \href{https://doi.org/10.1103/PhysRevD.96.115029}{\emph{Phys. Rev. D}
  {\bfseries 96} (2017) 115029}
  [\href{https://arxiv.org/abs/1707.02460}{{\ttfamily 1707.02460}}].

\bibitem{Acuna:2021rbg}
J.T.~Acu\~na, P.~Stengel and P.~Ullio, \emph{{A Minimal Dark Matter Model for
  Muon g-2 with Scalar Lepton Partners up to the TeV Scale}},
  \href{https://arxiv.org/abs/2112.08992}{{\ttfamily 2112.08992}}.

\bibitem{Kumar:2016cum}
J.~Kumar, P.~Sandick, F.~Teng and T.~Yamamoto, \emph{{Gamma-ray Signals from
  Dark Matter Annihilation Via Charged Mediators}},
  \href{https://doi.org/10.1103/PhysRevD.94.015022}{\emph{Phys. Rev. D}
  {\bfseries 94} (2016) 015022}
  [\href{https://arxiv.org/abs/1605.03224}{{\ttfamily 1605.03224}}].

\bibitem{Kelso:2014qja}
C.~Kelso, J.~Kumar, P.~Sandick and P.~Stengel, \emph{{Charged mediators in dark
  matter scattering with nuclei and the strangeness content of nucleons}},
  \href{https://doi.org/10.1103/PhysRevD.91.055028}{\emph{Phys. Rev. D}
  {\bfseries 91} (2015) 055028}
  [\href{https://arxiv.org/abs/1411.2634}{{\ttfamily 1411.2634}}].

\bibitem{Sandick:2016zut}
P.~Sandick, K.~Sinha and F.~Teng, \emph{{Simplified Dark Matter Models with
  Charged Mediators: Prospects for Direct Detection}},
  \href{https://doi.org/10.1007/JHEP10(2016)018}{\emph{JHEP} {\bfseries 10}
  (2016) 018} [\href{https://arxiv.org/abs/1608.00642}{{\ttfamily
  1608.00642}}].

\bibitem{Fukushima:2013efa}
K.~Fukushima and J.~Kumar, \emph{{Dipole Moment Bounds on Dark Matter
  Annihilation}}, \href{https://doi.org/10.1103/PhysRevD.88.056017}{\emph{Phys.
  Rev. D} {\bfseries 88} (2013) 056017}
  [\href{https://arxiv.org/abs/1307.7120}{{\ttfamily 1307.7120}}].

\bibitem{Dutta:2017nqv}
B.~Dutta, K.~Fantahun, A.~Fernando, T.~Ghosh, J.~Kumar, P.~Sandick et~al.,
  \emph{{Probing Squeezed Bino-Slepton Spectra with the Large Hadron
  Collider}}, \href{https://doi.org/10.1103/PhysRevD.96.075037}{\emph{Phys.
  Rev. D} {\bfseries 96} (2017) 075037}
  [\href{https://arxiv.org/abs/1706.05339}{{\ttfamily 1706.05339}}].

\bibitem{Baum:2020gjj}
S.~Baum, P.~Sandick and P.~Stengel, \emph{{Hunting for scalar lepton partners
  at future electron colliders}},
  \href{https://doi.org/10.1103/PhysRevD.102.015026}{\emph{Phys. Rev. D}
  {\bfseries 102} (2020) 015026}
  [\href{https://arxiv.org/abs/2004.02834}{{\ttfamily 2004.02834}}].

\bibitem{Pierce:2013rda}
A.~Pierce, N.R.~Shah and K.~Freese, \emph{{Neutralino Dark Matter with Light
  Staus}},  \href{https://arxiv.org/abs/1309.7351}{{\ttfamily 1309.7351}}.

\bibitem{XENON:2017vdw}
{\scshape XENON} collaboration, \emph{{First Dark Matter Search Results from
  the XENON1T Experiment}},
  \href{https://doi.org/10.1103/PhysRevLett.119.181301}{\emph{Phys. Rev. Lett.}
  {\bfseries 119} (2017) 181301}
  [\href{https://arxiv.org/abs/1705.06655}{{\ttfamily 1705.06655}}].

\bibitem{Mount:2017qzi}
B.J.~Mount et~al., \emph{{LUX-ZEPLIN (LZ) Technical Design Report}},
  \href{https://arxiv.org/abs/1703.09144}{{\ttfamily 1703.09144}}.

\bibitem{Dutta:2014jda}
B.~Dutta, T.~Ghosh, A.~Gurrola, W.~Johns, T.~Kamon, P.~Sheldon et~al.,
  \emph{{Probing Compressed Sleptons at the LHC using Vector Boson Fusion
  Processes}}, \href{https://doi.org/10.1103/PhysRevD.91.055025}{\emph{Phys.
  Rev. D} {\bfseries 91} (2015) 055025}
  [\href{https://arxiv.org/abs/1411.6043}{{\ttfamily 1411.6043}}].

\bibitem{Sandick:2016zeg}
P.~Sandick, K.~Sinha and T.~Yamamoto, \emph{{Black Holes, Dark Matter Spikes,
  and Constraints on Simplified Models with $t$-Channel Mediators}},
  \href{https://doi.org/10.1103/PhysRevD.98.035004}{\emph{Phys. Rev. D}
  {\bfseries 98} (2018) 035004}
  [\href{https://arxiv.org/abs/1701.00067}{{\ttfamily 1701.00067}}].

\bibitem{Kumar:2013iva}
J.~Kumar and D.~Marfatia, \emph{{Matrix element analyses of dark matter
  scattering and annihilation}},
  \href{https://doi.org/10.1103/PhysRevD.88.014035}{\emph{Phys. Rev. D}
  {\bfseries 88} (2013) 014035}
  [\href{https://arxiv.org/abs/1305.1611}{{\ttfamily 1305.1611}}].

\bibitem{Kumar:2016gxq}
J.~Kumar and C.~Light, \emph{{Connecting Dark Matter Annihilation to the Vertex
  Functions of Standard Model Fermions}},
  \href{https://doi.org/10.1088/1475-7516/2017/07/030}{\emph{JCAP} {\bfseries
  07} (2017) 030} [\href{https://arxiv.org/abs/1612.00773}{{\ttfamily
  1612.00773}}].

\bibitem{L3:2000bql}
{\scshape L3} collaboration, \emph{{Search for manifestations of new physics in
  fermion pair production at LEP}},
  \href{https://doi.org/10.1016/S0370-2693(00)00887-X}{\emph{Phys. Lett. B}
  {\bfseries 489} (2000) 81}
  [\href{https://arxiv.org/abs/hep-ex/0005028}{{\ttfamily hep-ex/0005028}}].

\bibitem{ATLAS:2019lff}
{\scshape ATLAS} collaboration, \emph{{Search for electroweak production of
  charginos and sleptons decaying into final states with two leptons and
  missing transverse momentum in $\sqrt{s}=13$ TeV $pp$ collisions using the
  ATLAS detector}},
  \href{https://doi.org/10.1140/epjc/s10052-019-7594-6}{\emph{Eur. Phys. J. C}
  {\bfseries 80} (2020) 123}
  [\href{https://arxiv.org/abs/1908.08215}{{\ttfamily 1908.08215}}].

\bibitem{CMS:2020bfa}
{\scshape CMS} collaboration, \emph{{Search for supersymmetry in final states
  with two oppositely charged same-flavor leptons and missing transverse
  momentum in proton-proton collisions at $\sqrt{s} =$ 13 TeV}},
  \href{https://doi.org/10.1007/JHEP04(2021)123}{\emph{JHEP} {\bfseries 04}
  (2021) 123} [\href{https://arxiv.org/abs/2012.08600}{{\ttfamily
  2012.08600}}].

\bibitem{ATLAS:2014fka}
{\scshape ATLAS} collaboration, \emph{{Searches for heavy long-lived charged
  particles with the ATLAS detector in proton-proton collisions at $ \sqrt{s}=8
  $ TeV}}, \href{https://doi.org/10.1007/JHEP01(2015)068}{\emph{JHEP}
  {\bfseries 01} (2015) 068} [\href{https://arxiv.org/abs/1411.6795}{{\ttfamily
  1411.6795}}].

\bibitem{Feng:2015wqa}
J.L.~Feng, S.~Iwamoto, Y.~Shadmi and S.~Tarem, \emph{{Long-Lived Sleptons at
  the LHC and a 100 TeV Proton Collider}},
  \href{https://doi.org/10.1007/JHEP12(2015)166}{\emph{JHEP} {\bfseries 12}
  (2015) 166} [\href{https://arxiv.org/abs/1505.02996}{{\ttfamily
  1505.02996}}].

\bibitem{ATLAS:2020wjh}
{\scshape ATLAS} collaboration, \emph{{Search for Displaced Leptons in
  $\sqrt{s} = 13$ TeV $pp$ Collisions with the ATLAS Detector}},
  \href{https://doi.org/10.1103/PhysRevLett.127.051802}{\emph{Phys. Rev. Lett.}
  {\bfseries 127} (2021) 051802}
  [\href{https://arxiv.org/abs/2011.07812}{{\ttfamily 2011.07812}}].

\bibitem{CMS:2019ybf}
{\scshape CMS} collaboration, \emph{{Searches for physics beyond the standard
  model with the $M_\mathrm{T2}$ variable in hadronic final states with and
  without disappearing tracks in proton-proton collisions at $\sqrt{s}=$ 13
  TeV}}, \href{https://doi.org/10.1140/epjc/s10052-019-7493-x}{\emph{Eur. Phys.
  J. C} {\bfseries 80} (2020) 3}
  [\href{https://arxiv.org/abs/1909.03460}{{\ttfamily 1909.03460}}].

\bibitem{Han:2014aea}
Z.~Han and Y.~Liu, \emph{{MT2 to the Rescue -- Searching for Sleptons in
  Compressed Spectra at the LHC}},
  \href{https://doi.org/10.1103/PhysRevD.92.015010}{\emph{Phys. Rev. D}
  {\bfseries 92} (2015) 015010}
  [\href{https://arxiv.org/abs/1412.0618}{{\ttfamily 1412.0618}}].

\bibitem{Barr:2015eva}
A.~Barr and J.~Scoville, \emph{{A boost for the EW SUSY hunt: monojet-like
  search for compressed sleptons at LHC14 with 100 fb$^{−1}$}},
  \href{https://doi.org/10.1007/JHEP04(2015)147}{\emph{JHEP} {\bfseries 04}
  (2015) 147} [\href{https://arxiv.org/abs/1501.02511}{{\ttfamily
  1501.02511}}].

\bibitem{ATLAS:2019lng}
{\scshape ATLAS} collaboration, \emph{{Searches for electroweak production of
  supersymmetric particles with compressed mass spectra in $\sqrt{s}=$ 13 TeV
  $pp$ collisions with the ATLAS detector}},
  \href{https://doi.org/10.1103/PhysRevD.101.052005}{\emph{Phys. Rev. D}
  {\bfseries 101} (2020) 052005}
  [\href{https://arxiv.org/abs/1911.12606}{{\ttfamily 1911.12606}}].

\bibitem{CMS:2019san}
{\scshape CMS} collaboration, \emph{{Search for supersymmetry with a compressed
  mass spectrum in the vector boson fusion topology with 1-lepton and 0-lepton
  final states in proton-proton collisions at $\sqrt{s}=$ 13 TeV}},
  \href{https://doi.org/10.1007/JHEP08(2019)150}{\emph{JHEP} {\bfseries 08}
  (2019) 150} [\href{https://arxiv.org/abs/1905.13059}{{\ttfamily
  1905.13059}}].

\bibitem{Alves:2017ued}
A.~Alves, T.~Ghosh and K.~Sinha, \emph{{Can We Discover Double Higgs Production
  at the LHC?}}, \href{https://doi.org/10.1103/PhysRevD.96.035022}{\emph{Phys.
  Rev. D} {\bfseries 96} (2017) 035022}
  [\href{https://arxiv.org/abs/1704.07395}{{\ttfamily 1704.07395}}].

\bibitem{ATLAS:2021kxv}
{\scshape ATLAS} collaboration, \emph{{Search for new phenomena in events with
  an energetic jet and missing transverse momentum in $pp$ collisions at $\sqrt
  {s}$ =13 TeV with the ATLAS detector}},
  \href{https://doi.org/10.1103/PhysRevD.103.112006}{\emph{Phys. Rev. D}
  {\bfseries 103} (2021) 112006}
  [\href{https://arxiv.org/abs/2102.10874}{{\ttfamily 2102.10874}}].

\bibitem{Chang:2010yk}
S.~Chang, J.~Liu, A.~Pierce, N.~Weiner and I.~Yavin, \emph{{CoGeNT
  Interpretations}},
  \href{https://doi.org/10.1088/1475-7516/2010/08/018}{\emph{JCAP} {\bfseries
  08} (2010) 018} [\href{https://arxiv.org/abs/1004.0697}{{\ttfamily
  1004.0697}}].

\bibitem{Feng:2011vu}
J.L.~Feng, J.~Kumar, D.~Marfatia and D.~Sanford, \emph{{Isospin-Violating Dark
  Matter}}, \href{https://doi.org/10.1016/j.physletb.2011.07.083}{\emph{Phys.
  Lett. B} {\bfseries 703} (2011) 124}
  [\href{https://arxiv.org/abs/1102.4331}{{\ttfamily 1102.4331}}].

\bibitem{Feng:2013vod}
J.L.~Feng, J.~Kumar and D.~Sanford, \emph{{Xenophobic Dark Matter}},
  \href{https://doi.org/10.1103/PhysRevD.88.015021}{\emph{Phys. Rev. D}
  {\bfseries 88} (2013) 015021}
  [\href{https://arxiv.org/abs/1306.2315}{{\ttfamily 1306.2315}}].

\bibitem{LUX:2013afz}
{\scshape LUX} collaboration, \emph{{First results from the LUX dark matter
  experiment at the Sanford Underground Research Facility}},
  \href{https://doi.org/10.1103/PhysRevLett.112.091303}{\emph{Phys. Rev. Lett.}
  {\bfseries 112} (2014) 091303}
  [\href{https://arxiv.org/abs/1310.8214}{{\ttfamily 1310.8214}}].

\bibitem{Cushman:2013zza}
P.~Cushman et~al., \emph{{Working Group Report: WIMP Dark Matter Direct
  Detection}},  in \emph{{Community Summer Study 2013}: {Snowmass on the
  Mississippi}}, 10, 2013 [\href{https://arxiv.org/abs/1310.8327}{{\ttfamily
  1310.8327}}].

\bibitem{PandaX-4T:2021bab}
{\scshape PandaX-4T} collaboration, \emph{{Dark Matter Search Results from the
  PandaX-4T Commissioning Run}},
  \href{https://doi.org/10.1103/PhysRevLett.127.261802}{\emph{Phys. Rev. Lett.}
  {\bfseries 127} (2021) 261802}
  [\href{https://arxiv.org/abs/2107.13438}{{\ttfamily 2107.13438}}].

\bibitem{Kopp:2014tsa}
J.~Kopp, L.~Michaels and J.~Smirnov, \emph{{Loopy Constraints on Leptophilic
  Dark Matter and Internal Bremsstrahlung}},
  \href{https://doi.org/10.1088/1475-7516/2014/04/022}{\emph{JCAP} {\bfseries
  04} (2014) 022} [\href{https://arxiv.org/abs/1401.6457}{{\ttfamily
  1401.6457}}].

\bibitem{DelNobile:2014eta}
E.~Del~Nobile, G.B.~Gelmini, P.~Gondolo and J.-H.~Huh, \emph{{Direct detection
  of Light Anapole and Magnetic Dipole DM}},
  \href{https://doi.org/10.1088/1475-7516/2014/06/002}{\emph{JCAP} {\bfseries
  06} (2014) 002} [\href{https://arxiv.org/abs/1401.4508}{{\ttfamily
  1401.4508}}].

\bibitem{Ho:2012br}
C.M.~Ho and R.J.~Scherrer, \emph{{Sterile Neutrinos and Light Dark Matter Save
  Each Other}}, \href{https://doi.org/10.1103/PhysRevD.87.065016}{\emph{Phys.
  Rev. D} {\bfseries 87} (2013) 065016}
  [\href{https://arxiv.org/abs/1212.1689}{{\ttfamily 1212.1689}}].

\bibitem{Gresham:2013mua}
M.I.~Gresham and K.M.~Zurek, \emph{{Light Dark Matter Anomalies After LUX}},
  \href{https://doi.org/10.1103/PhysRevD.89.016017}{\emph{Phys. Rev. D}
  {\bfseries 89} (2014) 016017}
  [\href{https://arxiv.org/abs/1311.2082}{{\ttfamily 1311.2082}}].

\bibitem{LUX:2015abn}
{\scshape LUX} collaboration, \emph{{Improved Limits on Scattering of Weakly
  Interacting Massive Particles from Reanalysis of 2013 LUX Data}},
  \href{https://doi.org/10.1103/PhysRevLett.116.161301}{\emph{Phys. Rev. Lett.}
  {\bfseries 116} (2016) 161301}
  [\href{https://arxiv.org/abs/1512.03506}{{\ttfamily 1512.03506}}].

\bibitem{LZ:2015kxe}
{\scshape LZ} collaboration, \emph{{LUX-ZEPLIN (LZ) Conceptual Design Report}},
   \href{https://arxiv.org/abs/1509.02910}{{\ttfamily 1509.02910}}.

\bibitem{LUX:2016ggv}
{\scshape LUX} collaboration, \emph{{Results from a search for dark matter in
  the complete LUX exposure}},
  \href{https://doi.org/10.1103/PhysRevLett.118.021303}{\emph{Phys. Rev. Lett.}
  {\bfseries 118} (2017) 021303}
  [\href{https://arxiv.org/abs/1608.07648}{{\ttfamily 1608.07648}}].

\bibitem{Fermi-LAT:2016uux}
{\scshape Fermi-LAT, DES} collaboration, \emph{{Searching for Dark Matter
  Annihilation in Recently Discovered Milky Way Satellites with Fermi-LAT}},
  \href{https://doi.org/10.3847/1538-4357/834/2/110}{\emph{Astrophys. J.}
  {\bfseries 834} (2017) 110}
  [\href{https://arxiv.org/abs/1611.03184}{{\ttfamily 1611.03184}}].

\bibitem{Geringer-Sameth:2011wse}
A.~Geringer-Sameth and S.M.~Koushiappas, \emph{{Exclusion of canonical WIMPs by
  the joint analysis of Milky Way dwarfs with Fermi}},
  \href{https://doi.org/10.1103/PhysRevLett.107.241303}{\emph{Phys. Rev. Lett.}
  {\bfseries 107} (2011) 241303}
  [\href{https://arxiv.org/abs/1108.2914}{{\ttfamily 1108.2914}}].

\bibitem{Fermi-LAT:2015kyq}
{\scshape Fermi-LAT} collaboration, \emph{{Updated search for spectral lines
  from Galactic dark matter interactions with pass 8 data from the Fermi Large
  Area Telescope}},
  \href{https://doi.org/10.1103/PhysRevD.91.122002}{\emph{Phys. Rev. D}
  {\bfseries 91} (2015) 122002}
  [\href{https://arxiv.org/abs/1506.00013}{{\ttfamily 1506.00013}}].

\bibitem{Boddy:2018qur}
K.~Boddy, J.~Kumar, D.~Marfatia and P.~Sandick, \emph{{Model-independent
  constraints on dark matter annihilation in dwarf spheroidal galaxies}},
  \href{https://doi.org/10.1103/PhysRevD.97.095031}{\emph{Phys. Rev. D}
  {\bfseries 97} (2018) 095031}
  [\href{https://arxiv.org/abs/1802.03826}{{\ttfamily 1802.03826}}].

\bibitem{Bergstrom:2013jra}
L.~Bergstrom, T.~Bringmann, I.~Cholis, D.~Hooper and C.~Weniger, \emph{{New
  Limits on Dark Matter Annihilation from AMS Cosmic Ray Positron Data}},
  \href{https://doi.org/10.1103/PhysRevLett.111.171101}{\emph{Phys. Rev. Lett.}
  {\bfseries 111} (2013) 171101}
  [\href{https://arxiv.org/abs/1306.3983}{{\ttfamily 1306.3983}}].

\bibitem{DiMauro:2015jxa}
M.~Di~Mauro, F.~Donato, N.~Fornengo and A.~Vittino, \emph{{Dark matter vs.
  astrophysics in the interpretation of AMS-02 electron and positron data}},
  \href{https://doi.org/10.1088/1475-7516/2016/05/031}{\emph{JCAP} {\bfseries
  05} (2016) 031} [\href{https://arxiv.org/abs/1507.07001}{{\ttfamily
  1507.07001}}].

\bibitem{Cheung:2009fc}
K.~Cheung, O.C.W.~Kong and J.S.~Lee, \emph{{Electric and anomalous magnetic
  dipole moments of the muon in the MSSM}},
  \href{https://doi.org/10.1088/1126-6708/2009/06/020}{\emph{JHEP} {\bfseries
  06} (2009) 020} [\href{https://arxiv.org/abs/0904.4352}{{\ttfamily
  0904.4352}}].

\bibitem{Fukushima:2011df}
K.~Fukushima, J.~Kumar and P.~Sandick, \emph{{Detection Prospects for Majorana
  Fermion WIMPless Dark Matter}},
  \href{https://doi.org/10.1103/PhysRevD.84.014020}{\emph{Phys. Rev. D}
  {\bfseries 84} (2011) 014020}
  [\href{https://arxiv.org/abs/1103.5068}{{\ttfamily 1103.5068}}].

\bibitem{Drees:1993bu}
M.~Drees and M.~Nojiri, \emph{{Neutralino - nucleon scattering revisited}},
  \href{https://doi.org/10.1103/PhysRevD.48.3483}{\emph{Phys. Rev. D}
  {\bfseries 48} (1993) 3483}
  [\href{https://arxiv.org/abs/hep-ph/9307208}{{\ttfamily hep-ph/9307208}}].

\end{thebibliography}\endgroup
